\documentclass[runningheads]{llncs}
\usepackage[T1]{fontenc}
\usepackage{graphicx}
\usepackage{url}
\usepackage{enumitem}
\usepackage{cite}
\usepackage{color}
\usepackage{appendix}
\usepackage[colorlinks=true, linkcolor=black, citecolor=black, urlcolor=black]{hyperref}
\usepackage[utf8]{inputenc}

\newcommand{\update}[1]{\textbf{}\textcolor{black}{#1}}

\begin{document}
\title{Lessons from Skill Development Programs - Livelihood College of Dhamtari}
%
%

\author{Arnab Paul Choudhury\inst{1}\orcidID{0000-0003-0739-8149} \and
Nihal Patel\inst{2,3}\orcidID{0009-0008-2158-7054}
}

\authorrunning{A. Paul Choudhury and N. Patel}

\institute{
Viksit Labs Foundation, Silchar, Assam, India, \email{arnabpchoudhury@viksitlabs.in} \email{arnabpc.oshin94@gmail.com} \url{https://www.viksitlabs.in}  \and
Indian Institute of Technology Guwahati, Guwahati, Assam, India \and
This work was done while the author was at the Indian Institute of Management Kozhikode, Mahatma Gandhi National Fellowship 
}

\maketitle              
\begin{abstract}

Skill training is crucial for enabling dignified livelihood opportunities. In India, various schemes and initiatives aim to provide skill training in different domains, with ICT and digital technologies playing a vital role. However, there is limited research on understanding on-ground capacities \& constraints and the use of digital tools in these programs. In this study, we look into the mobilization, counseling, and training stages of the 5-stage skill development process that also includes placement and tracking, adopted in Dhamtari's Livelihood College in Chhattisgarh, India, and other programs nationwide. Through the immersion/crystallization approach and mixed-method analysis including GIS mapping, video analysis of CCTV streams, quantitative analysis, and unstructured conversations with administrators, trainers, mobilizers, counselors, and nearby industry personnel for over a year, we identified three major challenges. A lack of inclusive and gendered access to skilling; a tedious manual counseling process with insufficient support staff; and inconsistent trainee attendance alongside sub-standard utilization of digital assets. Finally, we discuss, ways to improve access to skill training by leveraging Vocational Training Partners(VTPs), ways to improve the utilization of existing digital assets, and considerations for improving the counseling process. We conclude by summarizing that skill development programs currently lack institutional elements that enable effective information exchange between stakeholders, thereby creating information bottlenecks that result in inefficiencies, hindering the service delivery. In sum, our study informs the HCI and ICTD literature on the on-ground challenges and constraints faced by stakeholders and the role of technology in supporting such initiatives.

\end{abstract}

\section{Introduction}
The impact of COVID-19, the Russia-Ukraine war, and AI technology disruption have significantly influenced global health, geopolitical dynamics, and economic trends, resulting in varied outcomes for labor markets worldwide\update{\cite{Globalla24:online}.} Notably, low- and lower-middle-income countries continue to experience heightened unemployment rates, particularly among individuals with basic education and women \cite{mospigov41:online, Unemploy59:online}. Concurrently, real wages are declining due to an ongoing cost-of-living crisis, and changing worker expectations, such as a desire for better work-life balance, job security, career growth opportunities, and a healthy work environment, have emerged as prominent global issues, including in India \cite{RePEc:fas:journl:v:13:y:2023:i:2:p:4-28, Kunal2022EmployerAG, doi:10.1177/18479790221112548, GlobalEc91:online}.
\update{India boasts the youngest workforce globally, with a median age of 28, and 68\% of its population in the working age group (15-64) now has a demographic dividend \cite{Youthcan75, Populati31:online}. This demographic advantage historically fueled economic growth, translating into a larger workforce, rapid urbanization, and industrialization, attracting investments in infrastructure and human capital development leading to increased economic productivity, a rise of the middle class, and a greater purchasing power  \cite{HOSAN2022129858, Demograp14:online}. However, making effective use of this demographic dividend requires a robust Skill Development process that can adapt to people from diverse socio-economic backgrounds across the country \cite{BukuPoli36:online}. The situation is especially dire for blue-collar jobs, which are often less adaptable to disruptions, underscoring the need for targeted efforts to minimize shocks for workers in this section of jobs \cite{Technolo82:online}. In response, the Government of India has launched several Skill Development initiatives to enable effective skilling in various trades with a goal to provide meaningful, dignified livelihood and effectively utilize India’s demographic dividend for its continued growth.}
In particular, the Government of Chhattisgarh has initiated the "Livelihood College" program which aims to impart training in vocational courses and upskill youth from low and middle-income backgrounds in all districts of the state \cite{HomePage84:online}. Prior literature surrounding skill development, ICT and HCI largely focuses on the perspectives of the beneficiaries and the design \& development of tools and technologies to support their skilling. However, very little is known about the challenges faced by the authorities and stakeholders in such skill training centers and their interactions with technology. Hence in this study, we explore the challenges faced by stakeholders of a livelihood college in \textit{mobilization, counseling, and training} stages of the 5-stage skill development process adopted in the Livelihood Colleges, which also includes \textit{placement and tracking}. \update{We followed the "immersion/crystallization" approach to conducting a year-long study involving stakeholders of a Livelihood College such as trainers, administrators, mobilizers, counselors, and also nearby industry personnel, through unstructured conversations, participant observation, GIS mapping, quantitative and video analysis to understand,}

\textbf{RQ1:} \textit{What are the challenges and 
bottlenecks in the implementation of the 5-stage process in the skill development programs adopted by the state of Chhattisgarh through Livelihood College and throughout India under national schemes?}

\textbf{RQ2:} \textit{How are digital tools implemented, and utilized in Livelihood Colleges and other skill development programs in India?}

The qualitative inputs revealed a lack of inclusive mobilization and gendered access to skill training, challenges with the overall counseling procedure, limited human resources, and inconsistent trainee attendance/presence in classrooms. The qualitative inputs' on lack of inclusivity and challenges in counseling, were further investigated and triangulated quantitatively using data on home addresses, gender of the trainees, GIS mapping of the before mentioned, and analysis of publicly available data on district demographics. Finally, the inputs on inconsistent trainee attendance/presence were tested by comparing the biometric attendance data with video analysis of CCTV streams using a Yolov5 object-detection model which shows discrepancies in the implementation of biometric attendance collection. Following these findings, we discuss reasons for the lack of trainee representation from different parts of the districts and ways to improve access to skilling, how to improve the utilization of existing digital assets, and considerations \update{on improving counseling procedure}. The learnings from this study can be utilized not just within the state of Chhattisgarh but throughout the country in various other skill development initiatives. In sum, this work makes the following contributions to HCI and ICTD literature focusing on the design and implementation of skill development programs in India:
\begin{enumerate}[label=\alph*)]
    \item We performed a year-long qualitative study that provides insights into the challenges faced in the mobilization, counseling, and training process of the Livelihood College skill development program.
    \item We performed video analysis, quantitative analysis, and GIS mapping to investigate and triangulate the qualitative inputs received from stakeholders at the Livelihood College.
    \item We identified challenges concerning access to skill development, counseling procedures, \update{limited and sub-optimal use of digital assets, and inadequate} attendance/presence of trainees in the skill development programs.
    \item We discuss the role of ICT and digital technologies in \update{improving} access to skill training; improving utilization and implementation of digital assets and highlighting considerations \update{on improving the counseling process of skill development programs in India using technology interventions.
    \item We conclude by summarizing that the livelihood college initiative and skill development programs in general currently lack institutional elements that allow for robust communication exchange between stakeholders leading to inefficiencies, and sub-standard service delivery.}
\end{enumerate}


\section{Background and Related Work}
\subsection{Current Skill Ecosystem in India}
\update{The Skill India initiative was launched by the government in 2015 to train Indians using a result-oriented framework for industry-related jobs \cite{skilldev}. The primary objective was to train the youth to secure a better livelihood through a convergence of the overall skilling ecosystem by strengthening institutional mechanisms at both national and state levels, building quality trainers and assessors, establishing robust monitoring \& evaluation systems, and providing access to skill training opportunities. This umbrella initiative incorporates various schemes and programs, encompassing the journey from education to skill acquisition and ultimately securing a sustainable livelihood. It integrates multiple elements, such as short-term training, employment fairs (rozgar melas), recognition of prior learning, special projects, and guidelines for monitoring and placement, all within a unified platform \cite{SkillInd58:online}. The institutional mechanism to implement this mission consists Governing council at the apex level, a Steering Committee, and a Mission Directorate (along with an Executive Committee) as the executive arm of the Mission supported by the National Skill Development Agency (NSDA), National Skill Development Corporation (NSDC), and the Directorate General of Training (DGT) all of which lie under the Ministry of Skill Development and Entrepreneurship. At the State level, States are encouraged to create State Skill Development Missions (SSDM) along the lines of National Skill Development Missions with a Steering Committee and Mission Directorate\cite{skilldevmission}.}

\subsection{Initiatives by the Government of Chhattisgarh:}
Similarly, the Chhattisgarh government has passed the Chhattisgarh Right of Youth to Skill Development Act 2013 to secure the right to opportunities for skill development to every person between the age of 15 and 45 years residing in Chhattisgarh, in any vocation of choice consistent with eligibility and aptitude \cite{chhattis96:online, skillgapnsdc}. 
\update{Contemporaneously, the “Livelihood College” initiative was piloted in the Dantewada district—a region affected by tribal and Naxal violence since 2011, and has been replicated in 28 districts by 2023 \cite{HomePage84:online}.
These colleges focus on two main areas: emerging market trends like hospitality and industrial stitching, etc., and local skill deficits in trades such as plumbing, electrical work, solar panel installation, mobile repair, and many more. 
Further, the livelihood colleges are required to ensure training facilities, lab equipment, training regularity, monitor \& evaluation through CCTV, prevent unethical practices by the administration, and verify trainer qualifications through Training of Trainers (ToT) Certification\cite{Monitori50:online, PMKVYpdf14:online}.} The Skill development process undertaken by different arms of the Government usually adopts some common practices \cite{PMKVYGui27:online, SkillSaa82:online}. The 5 major steps that are generally adopted are detailed below:

\begin{itemize}
    \item {\textbf{Mobilization:}}
    \update{In this initial stage, mobilizers visit gram panchayats - a unit of local self-governance - to identify potential beneficiaries. Diverse methods, such as door-to-door mobilization, panchayat-level engagement, and local newspaper advertisements, are employed to raise awareness among youth and pique their interest.}
    \item{\textbf{Counseling:}}
    \update{In the Counseling stage, appointed counselors guide potential trainees on the schemes and the benefits of skill certification. The mobilizers and counselors are responsible for site selection accommodating a minimum of 30 candidates per session and ensuring accessibility and comfort. After counseling, batches of up to 30 trainees per trade/course are formed, and skill training commences.}

    \item{\textbf{Training:}}
    \update{The Training phase involves providing skill training in various trades chosen by the candidates.
    The training duration is four hours per day, and biometric attendance is collected to monitor attendance. 
    Candidates also participate in a minimum seven-day on-the-job training coordinated by the Livelihood College with potential employers. This experience is further followed by theoretical training, preparing candidates for deployment in the market.} 
    
    \item{\textbf{Placement:}}
    \update{Finally, the training institute takes the initiative to organize job fairs upon the successful completion of the course primarily for courses where employment opportunities are sought. Trainees can also secure loans from banks to start their ventures by presenting their certificates where they are assisted in business planning and obtaining bank approvals.}
    
    \item{\textbf{Tracking:}}
    \update{During this phase, the livelihood college administration monitors the placed trainees for a duration of 12 months supporting them through homesickness or harassment and assessing their employment status.}
\end{itemize}

\subsection{\update{Impact and effectiveness of skill development}}
\update{Studies have also been conducted to understand the effectiveness of skill development programs in India.
Studies show that in addition to increasing the economic status of trainees, it brings about apathy toward migration \cite{ghosh2022skilling}. Skill training has also been found to improve job performance, productivity, and promote self-employment among unemployed youth \cite{agrawal2020demographic, sinha2022impact}. A perception study in Karnataka found that training had medium effectiveness although trainees perceived that it helped them improve their knowledge and skills \cite{Kumar_Nain_Singh_Kumbhare_Parsad_Kumar_2021}. However, some studies also show that training does enhance job market prospects however other labor market forces, like caste-based discriminations, undo the positive effects \cite{chakravorty2019skills}. One study that examines factors that facilitate and hinder women's enrolment in skill development programs found that women who joined and completed courses could better cope with domestic economic challenges \cite{Sheshadri_Pradeep_Chandran_2021}. Studies also find that training semi-literate youth from disadvantaged rural backgrounds does improve "soft skills" but then places them in undesirable low-paid urban services leading them to quit in a few weeks and return home in search of other job opportunities \cite{doi:10.1080/01436597.2022.2077184}.}

Many more such studies have also been conducted that assess and examine the impact of such training programs which bring out the nuances in the outcomes, although most agree that training does lead to positive outcomes in knowledge and skill gain \cite{Swain2009impact, roy2018knowledge, agrawal2019impact, Tiwari_Malati_2020, Pandey_Gupta_Pandey_Singh_2017}.

\subsection{Skill Development and HCI} Prior literature surrounding skill development, vocational training, livelihood, labor, and HCI in India has primarily examined how digital technologies impact vocational workers. Work done to examine the perceptions and practices of vocational workers on automation found that technicians were unaware of the growth of automation but upon learning about it they expressed that they felt excluded by current technological platforms\cite{10.1145/3313831.3376674}. Another study that examined the experiences of workers on digital labor platforms acknowledges the evolving conditions surrounding skill-building among platform workers and highlights the need to support worker-centered skill-building \cite{10.1145/3596671.3597655}. Studies have also been conducted to design, develop, and evaluate technologies to support skilling using Augmented reality, and haptic simulation to make skilling more accessible \cite{10.1145/3380867.3426199, 6700139}. One study that details the design and measures efficacy of Technical Vocational Education and Training among low-literate rural users found that users perform better with ICT interventions \cite{10.1007/978-3-319-67684-5_1}. Earlier studies also attempted to design and deploy computerized vocational training courses on mobile-learning platforms delivered using automobile units \cite{6208644}. Some studies also focus on the need to update the skills of the gig economy amidst trends such as working from home, AI-enabled automation and the role of industry and government policies in supporting gig workers and employees prepare for the future of work \cite{gig_economy, doi:10.1177/26314541211064726}. Scholars have also designed applications to support migrant workers during the COVID-19 pandemic search for jobs and develop skills through videos and interactions through video \cite{10.1007/978-3-030-90238-4_2}. A study that looks into the PMKVY trainee perspectives concludes that even though skill training can provide opportunities for youth from working-class backgrounds, it is considered a last resort for trainees with poor academic records and is associated with lower esteem. \cite{doi:10.1080/0145935X.2024.2340551}.

Although multiple such studies exist in literature that hope to understand the perspectives of the beneficiaries and develop technologies to support their skilling, very little is known about the challenges and constraints the stakeholders such as trainers, administrators, counselors, and mobilizers face in implementing skill training programs and how technologies are used in supporting their work.

\subsection{Digitalization of Skill Development Programs in India}

The Government of India (GoI) has given a lot of thrust to digitalizing India on all fronts. Starting from mid-1990s, the GoI started multiple e-governance projects to provide citizen-centric services. This was further improved upon by the national level e-governance program called National e-Governance Plan (NeGP) initiated in the year 2006 \cite{Introduc1:online}. Finally, in the year 2015, the GoI launched the Digital India Mission which acted as an umbrella initiative for supporting multiple government schemes including the PMKVY which is part of the Skill India Mission and Livelihood College initiative\cite{DigitalI21:online}. The New Education Policy, of 2020 also highlights the need for integrating ICT and other digital tools in classrooms and calls for more interactive teaching \cite{NEPFinal2:online, Waoo_Waoo_2022}. Along similar lines, the livelihood college initiative also utilizes ICT technologies such as geo-tagging venues for counseling, biometric attendance, and CCTV installations for appropriate monitoring and evaluation \cite{SkillSaa82:online, PMKVYGui27:online, PMKVYpdf14:online}.


Although existing policies highlight the need to decrease the digital divide and utilize ICT to increase the efficacy of skill development and educational programs, very little is documented on how digital tools are implemented on the ground and the challenges \& concerns surrounding them from real-life implementations in skill development programs. Additionally, little to no work documents the service delivery of the Livelihood College initiative in Chhattisgarh, along with the impact and use of digital tools to augment the 5-stage process. Keeping in mind that a similar 5-stage process is adopted not just in the state of Chhattisgarh but also in the National level Skill Development programs in India, it becomes imperative to investigate how effective is the 5-step process and what more needs to be done to effectively deliver such skill development programs in different parts of the country. The learnings from this study can be used to identify key areas that need further improvement in service delivery and bring forth better outcomes for the beneficiaries. \update{To fill this critical gap, we indulged in site immersion at a representative Livelihood College and performed a mixed-methods analysis to draw our findings.} 




\section{Methodology}
As a part of the Chhattisgarh Right of Youth to Skill Development Act, the State Authority, CSSDA is mandated to conduct Research and Development in collaboration with various bodies to improve upon the service delivery of the skill development process \cite{chhattis96:online}. Hence, this work was done under due recognition and consent of the Office of the District Skill Development Authority as well as assent of the stakeholders and the authors were provided physical access to the facility \& data regarding various aspects of the Livelihood College during the study. \update{We followed the "immersion/crystallization" approach that guided our year-long study. The process was iterative involving multiple stages of data gathering and analysis while remaining reflexive throughout the process \cite{crabtree2023doing}.}

\subsection{Geographical area}
This study focussed on the Dhamtari district of Chhattisgarh, India. Dhamtari district is situated in the lower central part, fertile plains of Chhattisgarh region. Dhamtari, Nagri, and Kurud are included as Tehsils, and Dhamtari, Nagri, Kurud, and Magarlod are included as the district blocks. The district has a total population of 7,99,781 of which 6,62,443 are in its rural areas spread throughout 370 Gram Panchayats (GPs) - a unit of local self-governance, which are further combined to form 4 blocks namely Dhamtari, Kurud, Magarlod, and Nagri each containing 94, 108, 66, and 102 Gram Panchayats respectively \cite{Dhantari14:online}. Demographically, Dhamtari has a mixed population and the district is also home to sizeable tribal groups who account for 25.96\% of the district's population \cite{IndiaCen49:online}. The district is also endowed with natural wealth in abundance with Mahanadi as the principal river and diversity in flora and fauna majorly in the Nagri region. 
\subsection{\update{Data}}
\update{In our study, we performed a mixed-methods analysis involving qualitative \& quantitative analysis along with GIS mapping, CCTV video, and biometric data analysis. Table \ref{datasource} lists the type of data and their source that have been utilized in this study for analysis.}

\begin{table}[ht]
\centering
\caption{Data description.}\label{datasource}
\update{
\begin{tabular}{|l|l|} 
\hline
\textbf{Data} & \textbf{Source} \\
\hline
Unstructured Conversations& Primary \\
CCTV data & Primary \\
\raggedright  Biometric attendance & Primary (system generated)\\
\raggedright trainee details & Secondary (from Livelihood College) \\
\raggedright Physical Map of Dhamtari & Secondary (Publicly available - Govt. of Chhattisgarh) \\
\raggedright District demographics & Secondary (Publicly available - Govt. of Chhattisgarh) \\
\hline
\end{tabular}
}
\end{table}

\noindent \subsection{Qualitative analysis:}
\update{We started the study by gathering information through unstructured conversations primarily with the administrators, employees (clerks, mobilizers, counselors), trainers, and local industry personnel \& businesses. The details of the participants are mentioned in table \ref{participant}.} The administrators, employees (clerks, mobilizers, and counselors), and trainers were responsible for the proper functioning of the Livelihood College. They were also the first point of contact for the authors in the Livelihood college hence their perspectives formed the initial outlook on the college's functioning. Following this we also engaged with local industry personnel and businesses who are often the primary employers of the trainees who graduate from the Livelihood College. One of the authors in the study was physically present at the Livelihood College throughout the entire duration of the study to investigate and further improve the service delivery and functioning of the College. This physical on-ground interaction lasted for more than a year. The author had the assent, privilege, and free access to interact with the people in the Livelihood College. This prolonged exposure and interaction with the various stakeholders along with passive observations gave us a firm grasp on the current capacities of the program implementation as well as the constraints. \update{Given the year-long study with unstructured conversations, the questions evolved over time. Some high-level questions aligned to our RQs that guided our interactions are added to appendix \ref{appendix}}. However, the conversations were conducted largely to identify the bottlenecks in the 5-step process undertaken by skill development programs in India. These conversations allowed us to collect qualitative data that enabled us to identify and filter out some major challenges that needed further investigation.
\begin{table}
\caption{Participant Details.}\label{participant}
\centering
\update{
\begin{tabular}{|l|l|l|}
\hline
Participant profession &  Number & Gender\\
\hline
Administrators &  2 & All male\\
Clerks &  6 & All male\\
Trainers & 3 & All male\\
Trainers who also worked as mobilizers & 3 & All male\\
Counselors who also worked as mobilizers & 1 & Female\\
Mobilizers & 1 & Female\\
Business person (retail shop owner, rice mill owner, etc.) & More than 10 & All male\\
\hline
\end{tabular}
}
\end{table}

\noindent \subsection{Quantitative analysis and GIS mapping:}
\begin{figure}[htb]
\centering
\includegraphics[scale=0.6]{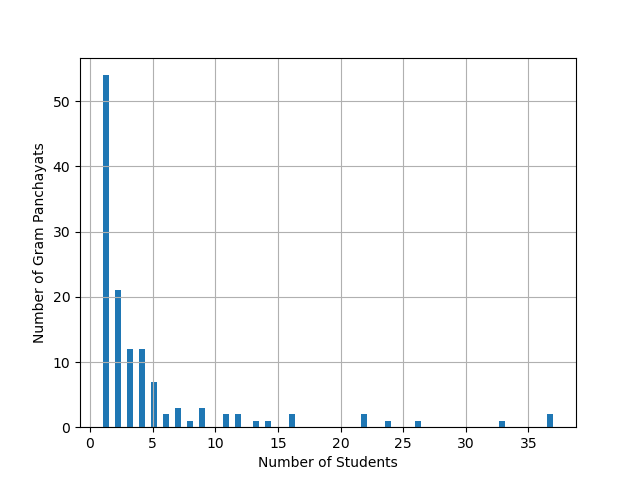}
\caption{Histogram showing number Gram Panchayats i.e., GPs (y-axis) contributing n-number of students/trainees (x-axis), (this excludes 81 trainees who were from cities/towns)} \label{histogram}
\end{figure}

Following the unstructured conversations, data on the trainee's attendance and their backgrounds were collected from the database of the Livelihood College. For this study, the latest data available at the time were for the year 2019-2020 where 670 trainees received training. 
The Financial year 2019-20 was the last year during which such data was collected as the activities of the college were disrupted since 2020 due to COVID-19. This data primarily included the trainee's name, their home address, the Gram Panchayat to which they belonged, district block, age, sex, and many more. This trainee data was first anonymized and de-identified to maintain the privacy of each trainee. Following this, the data was further filtered to remove information on those trainees who did not provide their place of residence or the Gram Panchayat to which they belonged. This resulted in 81 trainees' data being filtered out while the remaining 589 were then visualized into a histogram as can be seen in Figure \ref{histogram}. Each bar in Figure \ref{histogram} shows the number of Gram Panchayats (y-axis) that contributed n-number of trainees (x-axis).

\begin{figure}[htb]
\centering
\includegraphics[height=\textwidth]{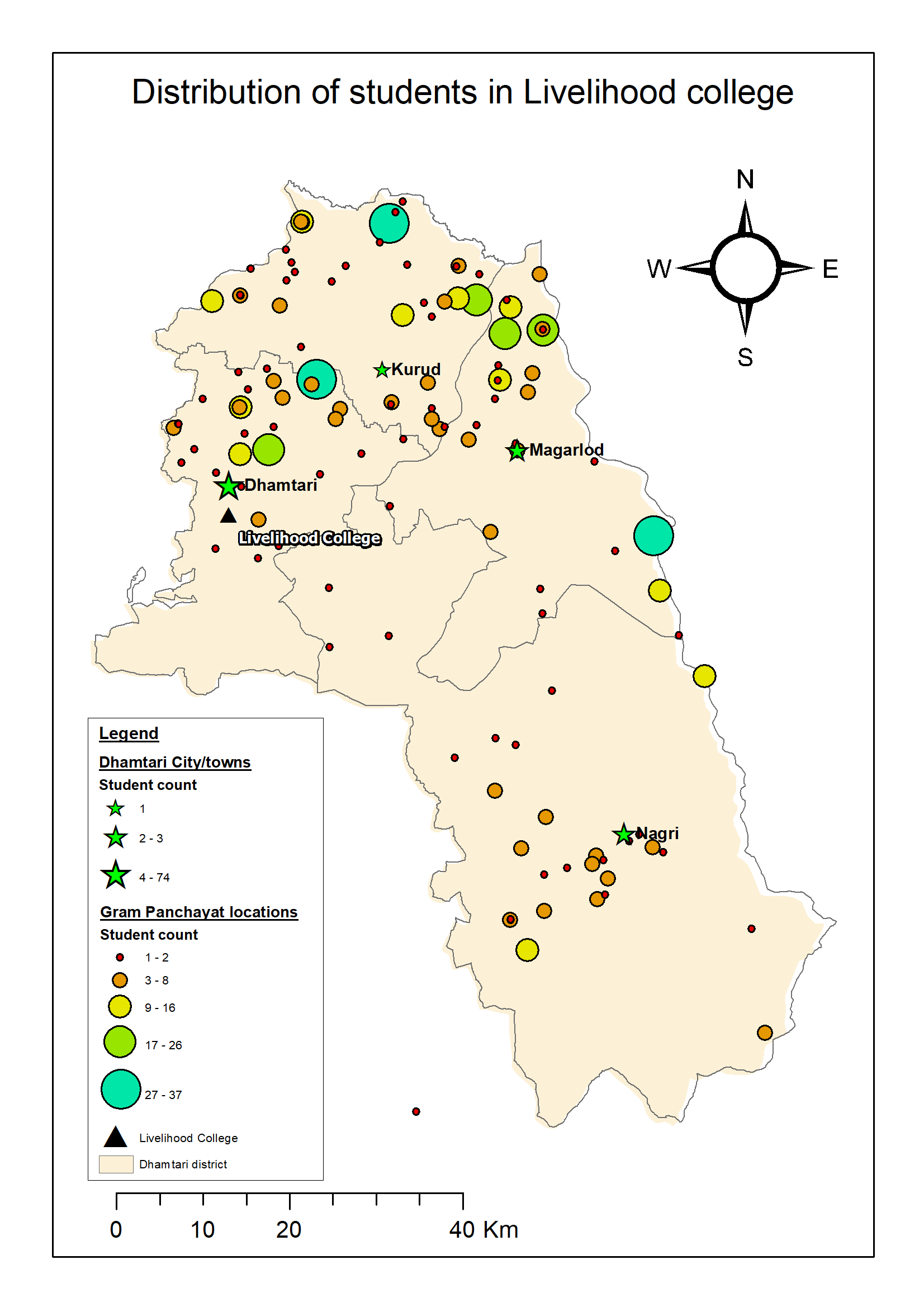}
\caption{Map showing the spatial distribution of students/trainees from different Gram Panchayat in Dhamtari} \label{dhamtarimap}
\end{figure}

\begin{figure}[htb]
\centering
\includegraphics[height=\textwidth]{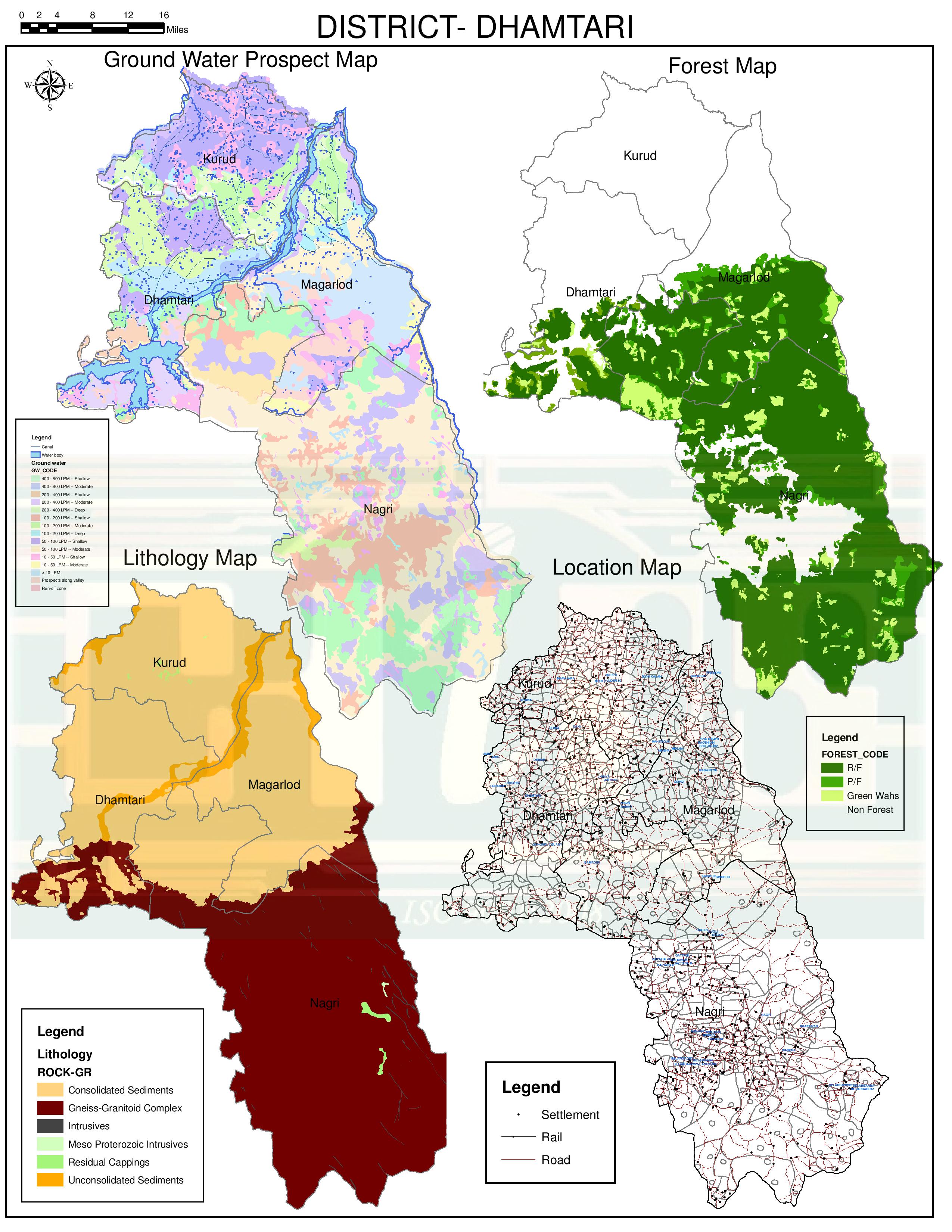}
\caption{Physical Map of Dhamtari district \cite{MapofDis47:online}} \label{dhamtariphysicalmap}
\end{figure}

\noindent It was later found that the 81 trainees were not from villages but were from towns/cities and hence the information on their Gram Panchayat was empty. \update{We also found that out of the 81, more than 70 were from the district headquarters of Dhamtari, the rest were from the major towns of the other three blocks.} Finally, we manually collected latitude and longitude values for each of the trainee address locations using Google Earth Pro software and utilized ArcMap software to create a GIS map which can be seen in Figure \ref{dhamtarimap}. The GIS map shows the spatial distribution of trainees along with the number of trainees studying in the Livelihood College from that particular location using graduated symbols. Finally, we also compare the information on the spatial distribution of trainees with the physical map of Dhamtari provided by the Government of Chhattisgarh that contains information on the location of Gram Panchayat's and settlements, forest map, lithology map, and groundwater prospects as can be seen in Figure \ref{dhamtariphysicalmap} \cite{MapofDis47:online}. Additionally, \update{we also compared} the publicly available data from the Government of Chhattisgarh's official website on the district's demographics with the data from the livelihood college which can be seen in Table \ref{tab1} \cite{BlockPan6:online}.

\noindent

\noindent

\begin{table}
\caption{Statistics on a block level}\label{tab1}
\begin{tabular}{|l|l|l|l|l|}
\hline
                                                    & \textbf{Dhamtari} & \textbf{Kurud} & \textbf{Magarlod} & \textbf{Nagri} \\
                                                    \hline
\textbf{Total number of GPs}                        & 94                & 108            & 66                & 102            \\
\textbf{Number of GPs represented}                  & 32                & 42             & 29                & 32             \\
\textbf{\% of GPs represented from block} & 34\%              & 39\%           & 44\%              & 31\%           \\
\textbf{Total Population of   GPs}                  & 183373            & 197723         & 115150            & 166197         \\
\textbf{Average number   person / GP}               & 1950.78          & 1830.77       & 1744.70          & 1629.38       \\
\hline
\textbf{Number of trainees   (GP only)}             & 133               & 169            & 172               & 115            \\
\textbf{\% of total trainees   (GP only)}           & 23\%              & 29\%           & 29\%              & 20\%           \\
\textbf{Number of trainees   (GP+city/town)}        & 207               & 170            & 175               & 118            \\
\textbf{\% of total trainees   (GP+city/town)}      & 31\%              & 25\%           & 26\%              & 18\%           \\
\hline
\textbf{Male-to-female   ratio (GP only)}           & 0.40              & 0.47           & 1.61              & 1.02           \\
\textbf{Male-to-female   ratio (GP+city/town)}     & 0.52              & 0.48           & 1.62              & 1.07           \\
\hline
\end{tabular}
\end{table}
\noindent


\subsection{Attendance analysis using CCTV and biometric data:}
\begin{figure}[htb]
  \centering
  \includegraphics[width=\linewidth]{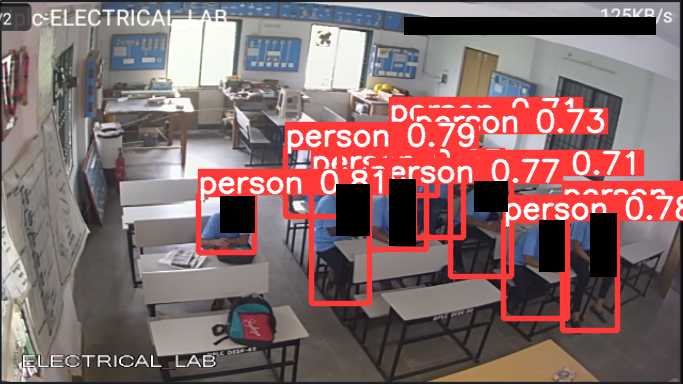}
  \caption{Representative image for person detection in class}\label{repimage}
\end{figure}

\begin{figure}[htb]
  \centering
  \includegraphics[width=0.9\linewidth]{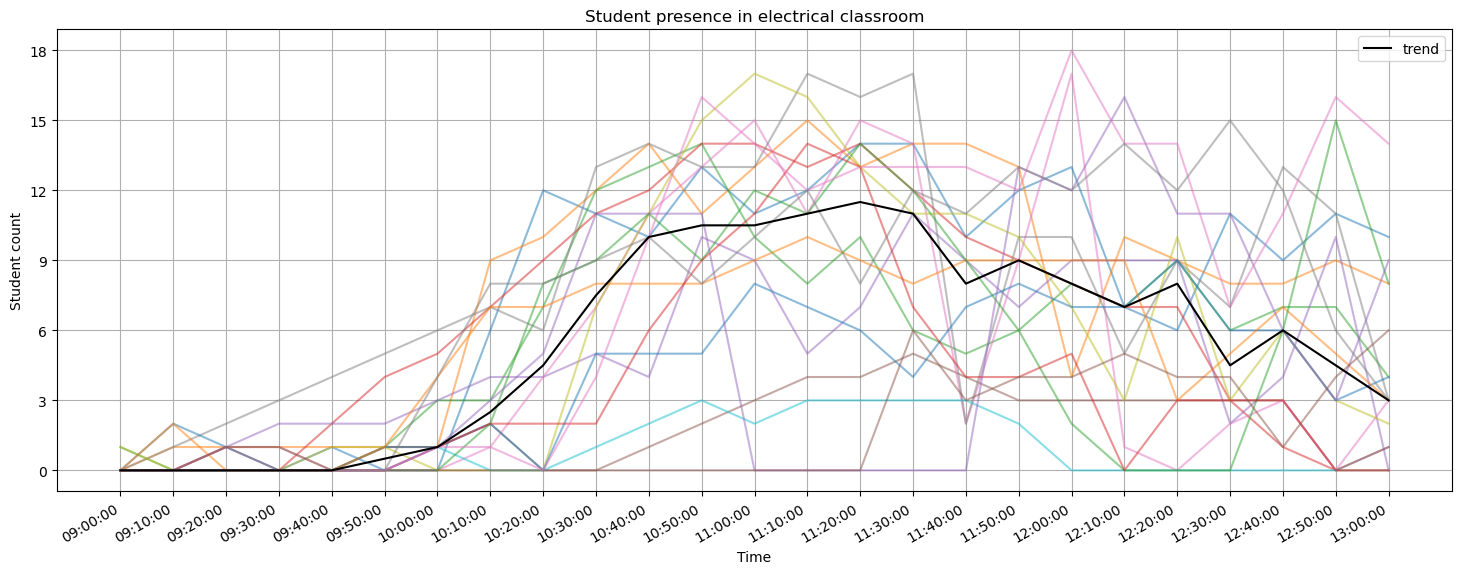}
  \caption{ Student/trainee presence in electrical classrooms, x-axis: time, y-axis: trainee count}\label{elestudpresence}
\end{figure}

\begin{figure}[htb]
  \centering
  \includegraphics[width=0.9\linewidth]{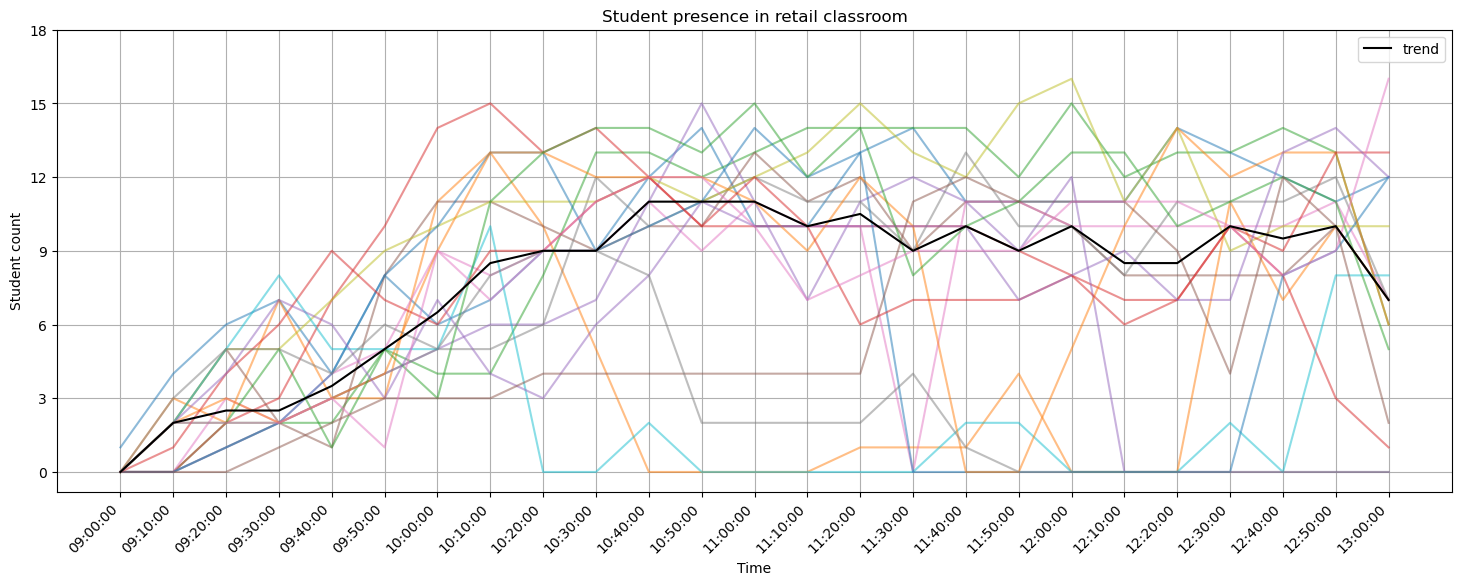}
  \caption{ Student/trainee presence in retail classrooms, x-axis: time, y-axis: trainee count}\label{retstudpresence}
\end{figure}

\begin{figure}[htb]
  \centering
  \includegraphics[width=0.8\linewidth]{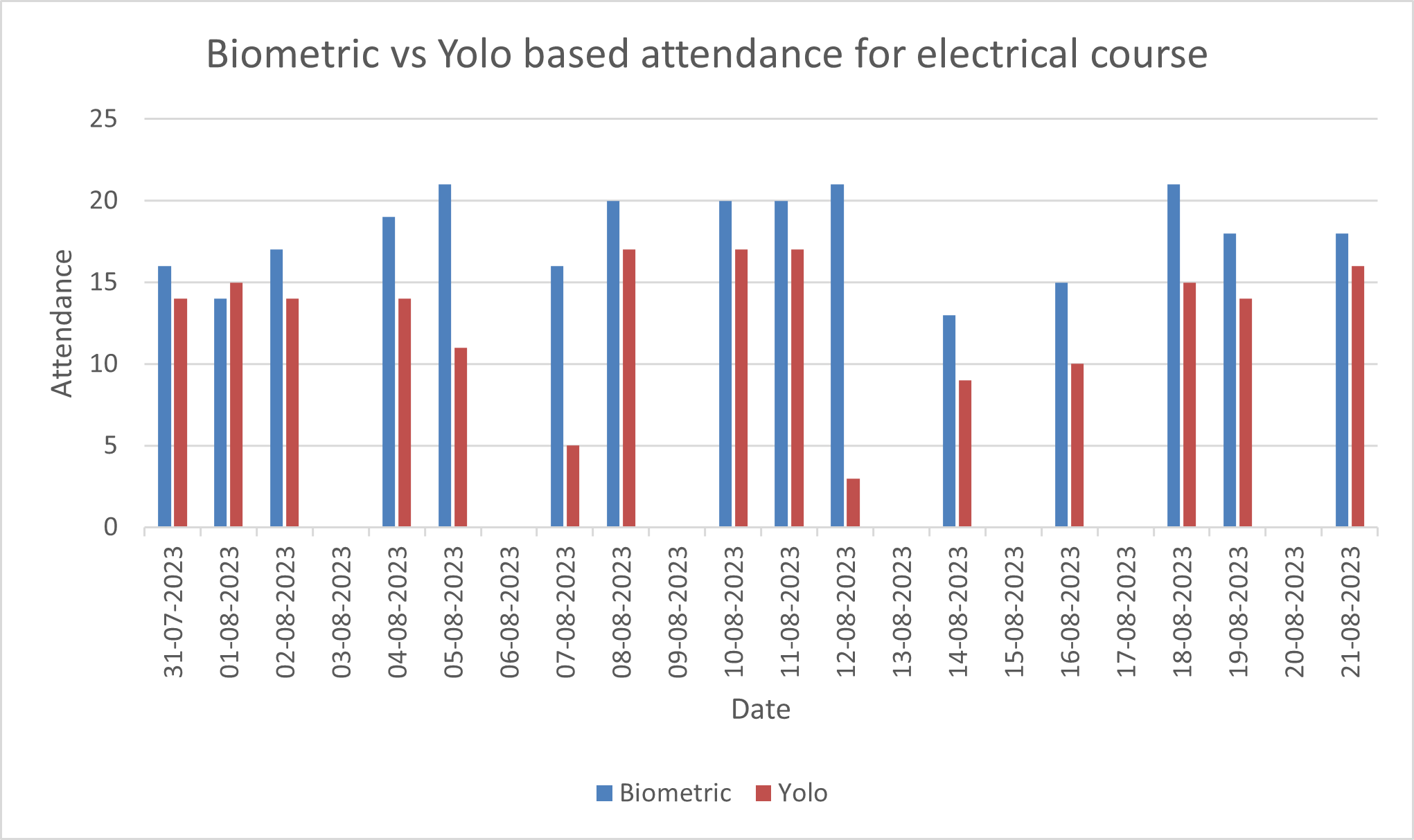}
  \caption{Biometric vs Yolo-based attendance for electrical course}\label{eleattndis}
\end{figure}

\begin{figure}[htb]
  \centering
  \includegraphics[width=0.8\linewidth]{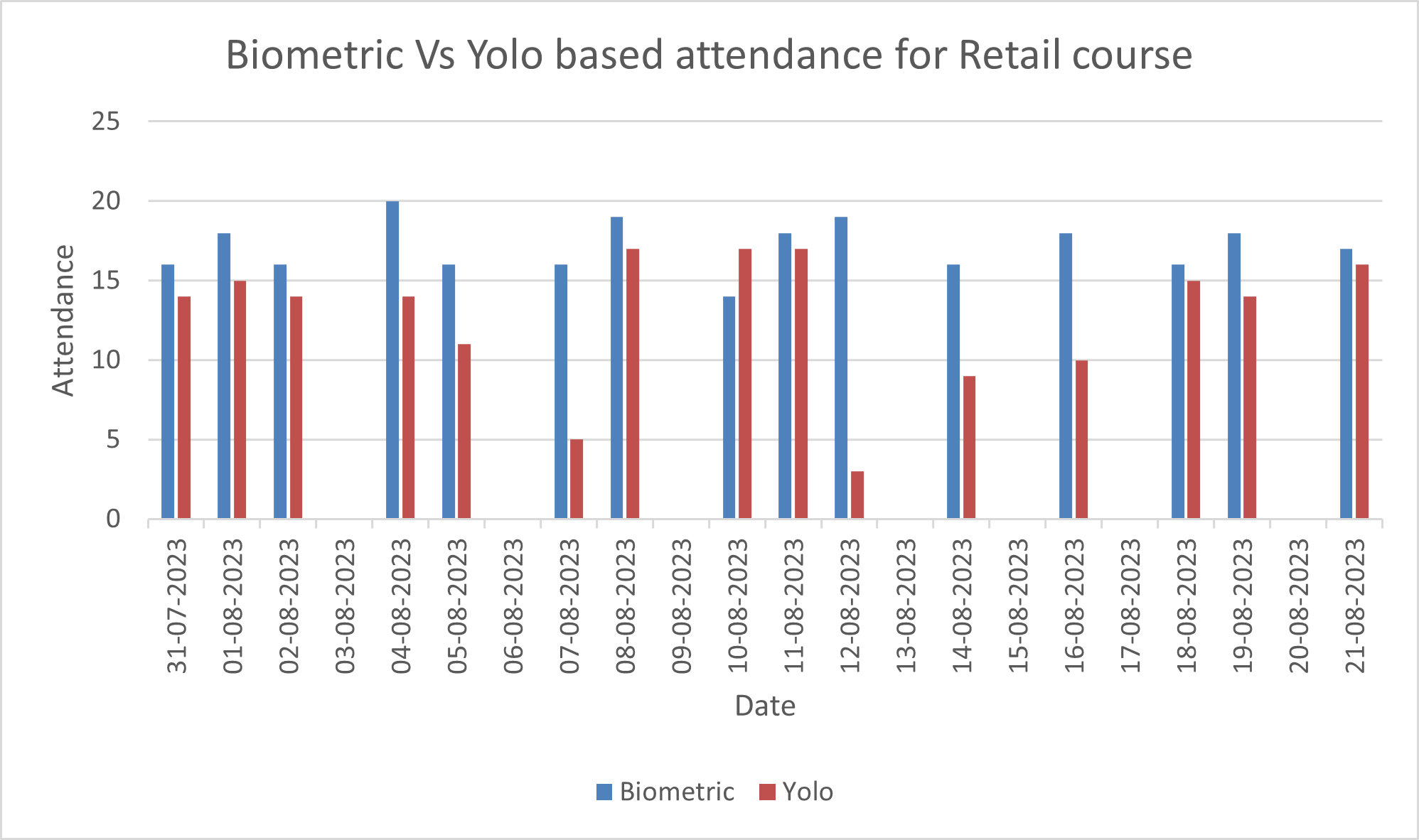}
  \caption{Biometric vs Yolo-based attendance for retail course}\label{retattndis}
\end{figure}

\noindent The CCTV cameras installed at the Livelihood College were also analyzed by taking snapshots at an interval of 10 mins every working day for 1 month from 09:00 am to 01:00 pm which are the usual working hours for our targeted courses i.e., Electrical and Retail. During the time of this study, the livelihood college had 4 running courses namely, Electrical, Retail, Computer hardware and software, and Sewing. However, upon inspection, it was found that the CCTV placement of all the courses except electrical and retail had some errors i.e. either the entire class was not visible, or the angles were incorrect. It was also noted that the CCTVs are usually unutilized assets and were seldom used for any productive purposes. In two of the CCTV cameras, the video feed was obstructed by spider webs and we had to manually clean them before attempting to utilize the video data. As a result, we performed video analysis only on Electrical and Retail courses. Since direct access to the video streams was not readily available through the CCTV application, we used the Bluestacks emulator to run the CCTV application on a desktop. A python script was written which would then take images at every 10-minute interval starting from 9:00 am to 01:00 pm. Consecutive courses were mapped onto the same screen and hence the images were further cropped to get the final set of images. In total 25 images were captured per day for each course and the same was repeated for 18 working days in the month. Finally, this resulted in a total of 900 images, 450 for each course. Along with taking the snapshots, the date and time at which they were captured were retained in their filenames in snake casing so as to ensure data integrity. Person detection was then performed using an open-source Object detection model, YOLOv5 with the pre-trained model checkpoint “yolov5x6u.pt” with the default parameters and threshold values. We used this pre-trained checkpoint as during this study it had one of the highest Mean Average Precision 50-95 (mAP50-95) values of 56.8\% among all known real-time object detection models \cite{YOLOv5Ul97:online}. The entire process of prediction and inference from the images was conducted on a Google Colab notebook. To further ensure that the model results were not erroneous, we randomly selected a subset of images from the 900 and manually checked for the correctness of the outputs which in most cases resulted in a correct count or at most had an error count of 1 or 2 in a handful of cases due to trainees overlapping each other while moving around to get inside or move out of the class which is not the usual practice during training sessions as trainees are expected to sit separately in their appropriate places which ensures no overlaps in the video feed. A reference image of a particular course is shown in Figure \ref{repimage} however the faces of individual trainees have been blacked out manually to retain their privacy. The predicted outputs from each snapshot were then collated together to generate time series data showing the number of trainees present in the classroom throughout the day. The same method was applied for all the other working days and both the courses respectively to create master data for the two courses with days as column headers and time as index with numbers in each cell indicating the trainee count at a particular date and time. Upon analyzing this table, we could identify some NULL values in cells which resulted from images that had blank screens primarily due to network issues. All such NULL values were removed and were filled by linearly interpolation. In addition, for the retail course, the lab had two dummy models placed within the premises of the class which were used as part of the training. The YOLOv5 model consistently identified these two dummies as persons, hence we corrected this error by negating 2 from all the predictions. Once the data was cleaned, a median count of trainees present at any given time was calculated for these 18 days. Median was used as a measure so as to limit the impact of outliers in the data. This was repeated for both courses respectively. The resulting median value time series is interpreted as the trend of trainees present in the classrooms at any given time during that month. Finally, the time series for each course was visualized using a line graph as can be seen in Figure \ref{elestudpresence} and Figure \ref{retstudpresence}. We also utilized the biometric data used by the trainees in the Livelihood College to register their attendance. Biometric data was available for 15 days as opposed to the 18 that was generated using the Yolov5 object detection model. Hence, only, 15 days of pairwise data was used for comparison between Biometric data and the output from the object detection model. The trainees typically register their attendance twice using the biometric attendance system using their fingerprints when they first reach the Livelihood College and when they move out. Since the biometric data generates a single value for each day representing the number of trainees present in class, we took the maximum number of trainees present at any given time found using Yolo to compare with the biometric data. The bar chart comparing the two can be seen in Figure \ref{eleattndis} and Figure \ref{retattndis}.

\noindent

\noindent

\noindent

\noindent

\noindent \subsection{\update{Positionality and reflexivity}}
\update{Our goal is to improve skill training service delivery such that it is inclusive, relevant, high-quality, and effective in generating meaningful livelihoods for beneficiaries. All authors are of Indian origin having experience working with low-middle-income and marginalized groups. One author, fluent in Chhattisgarhi, interacted with the participants and beneficiaries for over a year. While our study participants aren’t necessarily marginalized and we didn’t exert significant influence or power, we acknowledge our privileged access to information, remain mindful of our assumptions \& biases, and strive for reflexivity throughout the research process centering the needs of the beneficiaries and participants.}
\section{Findings}
\update{Our analysis revealed multiple challenges and constraints faced in the service delivery of the skill development program.} In the below section, we highlight our major findings:

\subsection{Lack of inclusive mobilization}
An initial mapping of a limited number of trainee data revealed some level of selective recruitment based on geography, in the livelihood college. Upon questioning the mobilizers and counselors on the same, they reported how limited resources disincentivized them to reach out to far-flung areas of the districts. In particular, they mentioned how reaching out to far-flung areas would not reap any benefits as there was no operational accommodation facility for housing trainees from these areas. They also raised concerns about the lack of better transportation and arrangements for reaching out to trainees from hard-to-reach interior regions of the district who could not travel to the livelihood college daily from their houses.
This was further investigated using the data available on trainees’ backgrounds for the year 2019-2020 which had information on the geographical location of the Gram Panchayats and communities to which they belong. The analysis showed that only 130 Gram panchayats are receiving the benefits of the courses being imparted at the Livelihood College out of the total 370 Gram Panchayats in Dhamtari leaving the remaining 240 Gram Panchayats completely untouched. Even among those 130 Gram Panchayats, Figure \ref{histogram} shows that only a few Gram Panchayats are contributing a substantial number of trainees. Most Gram Panchayats only have 1-4 trainee representatives. Similarly, from Figure \ref{dhamtarimap} we can see that most of the trainees are clustered around the major cities and towns of the district with the highest representation from the upper regions of the Dhamtari and Kurud blocks. Looking at Table 1, it becomes clear that in almost all blocks the percentage of Gram Panchayats being represented is less than 50\% in all with Nagri being the least at 31\%. If we consider the number of trainees from Gram Panchayats of different blocks we do not see a stark difference in representation though Kurud and Magarlod contribute slightly higher number of trainees. However, when we consider the numbers along with the city of Dhamtari and other major district towns we see that Nagri contributes 18\% of the total trainees which is almost half of Dhamtari contributing 31\%. Taking a look into the male-to-female ratio, we find that in the blocks of Kurud and Dhamtari, females seem to be attending the classes in more numbers as compared to trainees from Magarlod where males are more in number.

\subsection{Tedious counseling procedure}
The counselors and mobilizers also reported that the current counseling process was manual, time-consuming, and did not guarantee a substantial turnover rate of trainees, which also contributed to the final selection process being less inclusive. They mentioned that the current counseling process is a multi-stage process where first, the counselors and mobilizers reach out to selected villages and mobilize prospective trainees to educate them about the livelihood college. Here, they are supported by organizations called Vocational Training Partners (VTPs), agencies that support the Livelihood College in reaching out to trainees from hard-to-reach areas. VTPs also conduct training but with highly limited capacity as they are heavily resource-constrained. The mobilizers take support from these VTPs who have social capital in the region, arrange for the venue, and mobilize candidates for the awareness programs. Next, the candidates are invited to the livelihood college where they are provided in-depth information about the benefits of the course. In reality, most of the candidates do not turn up at the livelihood college, and only a fraction visit. Following this, even among the candidates who visit the Livelihood College, only a few decide to join. In most cases, the number of candidates who show interest is less than 30, which is the suggested number of trainees required to start a course. On rare occasions, when there are more than 30 candidates, the counselor has the freedom to select the ones who show the most promise and withhold the rest for the next batch based on their subjective analysis. The counselors and mobilizers noted that these hurdles primarily arise as it is challenging to ascertain the right fitment of a particular trainee in a course, identify the comprehension levels, and interests of trainees, and provide appropriate personalized guidance. Rather, at present they have to counsel people in masses at one go during the physical village visits or when some trainees visit the livelihood college in mass, neither of which allows them to understand the constraints of the trainees or provide individualized feedback which might encourage or motivate them to join the training. 

\subsubsection{Limited human resources impacting trainers}
A major reason for the counseling procedure being tedious was the lack of required human resources. The livelihood college primarily requires the following of human resources for its functioning, administrators, clerks, counselors, mobilizers, and trainers.
From our conversations with the trainers, it became evident that the ranks of the mobilizers, counselors, and clerks were not fulfilled appropriately and very often there were vacant positions. This essentially meant that the trainers had to tend to the extra work and perform mobilization, and counseling for the trainees, thereby putting a lot of burden on them and reducing their capacity to perform their actual job roles and responsibilities which included sorting course structures, making lesson plans, evaluating trainees answer sheets, and many more. Upon asking the stakeholders, what might be the reasons for the continual vacant positions, they unanimously agreed that the job remuneration and career growth were not lucrative enough for someone to engage in the position for a longer term.

\subsection{Inadequate trainee presence in classrooms}
A few of the stakeholders also reported absenteeism in classes despite registering attendance using biometric data. We looked at the biometric attendance data and found that using this data we could not conclude that trainees remain absent from classes as they were just point values in a day. Hence to investigate this, we analyzed the CCTV streams. As can be seen from Figure \ref{elestudpresence} and Figure \ref{retstudpresence}, most trainees arrive late at classes indicating that the classes are not starting as per the schedule and the time spent in classrooms by a majority of the trainees is almost half the actual allotted time for instructions. The situation in the retail course is perhaps slightly better where once the trainees are inside the classes they maintain their presence throughout the entire duration however for the electrical course it seems trainees not only arrive late but also leave early which is indicated by the declining trend in classroom presence from around 12:00 pm. Taking a look at Figure 5 and Figure 6 makes it clear that the maximum number of trainees who attend the classes also seems to be capped well below the total strength of the trainees found present through the Biometric data. On some days, the biometric data and the Yolo detected data are nearly close indicating that trainees do ensure to register their biometric attendance even if they arrive late and move out early. However, we also found a few days where the Yolo models detected very little presence as compared to the biometric data, and the difference is quite stark. 
\subsubsection{Method of registering biometric attendance.}
A few of the stakeholders we spoke to mentioned various reasons ranging from lack of interest to trainees from low-income backgrounds requiring to support their family business during college hours and many more as reasons for the inadequate trainee attendance and presence. Following this, we enquired the stakeholders about the biometric attendance registration procedure in depth. They reported that trainees utilize their Aadhar information -  an individual identification number acknowledged by the Government of India as proof of address and identity, to provide biometric attendance using fingerprints every working day\cite{WhatisAa54:online}. The system registers attendance as present for the day only if a trainee has punched in their information twice during the 24-hour cycle. Additionally, the system also captures the timestamp during which the two punches were made. However, the duration between the two punches is not taken into consideration for attendance registration. Hence, a trainee spending 4 hours in the class may show the same attendance as someone who has hardly spent an hour or none at all.
\subsubsection{Utilization of CCTV.}
While attempting to analyze the CCTV streams, it was noticed that many of the CCTVs were not placed at the correct angles meaning that the entire class was not visible from the camera. In addition, some also had cobwebs blocking the video feed which indicated that the video was not being utilized for any practical purposes or served any utility for the day-to-day operations of the livelihood college though one key reason for the installation of the CCTV was to ensure proper monitoring and evaluation. 

\section{Discussions}
Drawing on the findings, we now discuss their implications and possible ways to mitigate challenges in the below sections:
\subsection{\update{Improving access to skill training}}
One of the key findings was that mobilizers and counselors were not motivated to reach out to certain villages and Gram Panchayats because they were aware that they couldn't provide residential accommodation to trainees from such farm-flung areas. \update{If we look at the distribution of trainees in Figure \ref{dhamtarimap}, and the physical maps in Figure \ref{dhamtariphysicalmap}, we can see more clearly how the upper regions of the district which are relatively more urbanized and well-connected owing to their proximity to the state capital of Raipur, contribute more trainees as compared to the lower and forested regions which are more rural and the transportation system is not as developed. Similarly, we also find that more females are being trained from the upper blocks which might be because they could travel back and forth every day as opposed to female trainees from the blocks of Magarlod and Nagri where a female trainee may have to rent houses adding in a cost factor.} Additionally, existing social norms and gender roles concerning women in rural India would not make it easy for them to rent a house and stay independently, away from their homes \cite{https://doi.org/10.1111/ajae.12114, BALACHANDRAN2024100060}. A major concern however is that the regions that are empty in Figure \ref{dhamtarimap} are typically Naxal-affected forest areas as can be seen in Figure \ref{dhamtariphysicalmap}, with a significant population of people from marginalized groups who should have been among the beneficiaries of such a program. Although the counselors and mobilizers did not exhibit any bias against marginalized groups during our conversations, we did notice a subtle indifference arising from their current constraints, and the inability to make any strides in this direction is a cause for major concern.

To address this concern, one effective approach could be to enhance the role of Vocational Training Providers (VTPs). Recognizing that operationalizing residential accommodations or improving transportation infrastructure — both of which demand significant political commitment and bureaucratic support—might take considerable time, a more immediate solution lies in improving the capacities of the VTPs which are physically in close proximity to the different villages. Currently, VTPs are not capable of imparting such training, and arranging physical resources for each individual VTP may be cost-intensive, instead it may be prudent to effectively use ICT tools and advances in low-cost AR/VR technologies to provide trainees with similar experiences that they may receive in the Livelihood College \cite{10.1145/3332165.3347889, 10.1145/3313831.3376674, CHIANG2022107125}.

\subsection{Improving utilization of digital assets}
One of the key observations made during the course of the study was that the digital assets in place were either not utilized properly or were completely left unused. Even though policies mandate the implementation of digital technologies to ensure better service delivery, how these technologies are implemented and utilized is largely left to be decided ad-hoc. For e.g.: the way the attendance is registered from the biometric devices has multiple scopes of errors. At present, even if a trainee punches in twice consecutively within a short period of time even after class hours, attendance is considered to be accepted. It might be so that a person has classes from 09:00 am to 01:00 pm but they arrive at 12:30 pm, punch in twice, and leave the premises. We also find evidence of such practices when we compare the results of the Yolo-based attendance with the biometric attendance. It is essential to ensure that the attendance captured correctly reflects the presence of trainees in the class as all policy-related decisions are taken by authorities considering such data and a wrong understanding of the on-ground situation would lead to no improvement in the process. For e.g., to a policymaker/administrator who may have little information on the day-to-day activities of the college, looking at the current biometric data would give a false impression that trainees are attending classes without any hiccups when the reality is far from this. This also raises questions as to why the trainees may be bunking classes or not attending the entire duration. There may be a multitude of answers to this and the best source of this information is perhaps the trainees themselves who were not part of this study. However, questions such as these can only be raised and investigated only when technologies are implemented appropriately and with due care. Following this, we also noted that many of the CCTV cameras had improper camera angles and some had cobwebs obstructing the video feed indicating that the video was not being utilized for any monitoring or evaluation purposes. An easy solution to biometric attendance is to consider the timestamp when the trainees punch in twice, this should provide enough information to understand if a trainee truly has attended classes. In addition, if the video feed is analyzed using an offline object detection model using systems that are already present and capable of checking for the presence of trainees, this information can further augment the biometric attendance data. Finally, the video data can also be used for various other purposes such as identifying the concentration and stress levels of trainees which can then be used to provide personalized support \cite{7808130, 9610134, 9057967}. However, one must also take note that such solutions preserve the privacy of the trainees and don't have any adverse impact on trainees \cite{Preuveneers2021-ud, Wu2023-mw}.

\subsubsection{Need for retrofit, offline, edge solutions.}
One way to increase the utilization of these assets is through low-cost retrofit solutions rather than those that need fundamental changes either on the technology stack or from a management perspective. For example, at the Livelihood College, some level of video analysis could be provided by the application running the CCTV, however, this feature was not made an explicit requirement in policy and hence was not subscribed as this was a paid service. Also, making such intricate changes in policy would be a tedious and lengthy process requiring approvals on multiple levels of the government hierarchy. Rather, the flexibility to make implementation decisions on the ground given to the administrators should be utilized to implement retrofit solutions that do not tamper with the existing infrastructure or require lengthy approvals either due to monetary or management reasons. E.g., in our case, the stakeholders in the livelihood college did not have direct access to the raw video stream or were unaware of how to access it. Hence, we utilized a retrofit method to do the video analysis which was much more feasible to implement on the ground with limited resources than having to gain access to the live video data. In addition, livelihood colleges are usually located in small towns \& cities and have to deal with resource constraints including lack of a high bandwidth or stable internet connectivity. In such a situation, depending on cloud services is not an ideal solution, rather making use of edge-based solutions on moderate to low-compute devices such as mobile/smartphones, laptops, and desktops may be a better alternative.

\subsection{\update{Considerations on improving counseling procedure}}
\update{One key challenge mentioned by the participants was the tedious and manual nature of the counseling procedure that primarily involves identifying the right fitment of trainees, comprehension levels, and interests and finally advising trainees on a career path. The major reason is a lack of adequate human resources to reach out to the beneficiaries and provide individualized feedback. Here too, the VTPs may play a critical role in reaching out to the beneficiaries given their physical proximity and on-ground presence. However, it would not be wise to assume that the VTPs would be capable of providing counseling to the beneficiaries appropriately, rather, they may act as points of contact at which people can receive counseling through video conferencing on ubiquitous platforms like WhatsApp, Zoom, MS Teams, Gmeet, etc., from expert counselors. Though some studies show that online career counseling intervention increases trainees' level of career development, existing interventions largely focus on service providers and augmenting their conventional practices with little attention to the beneficiaries \cite{Pordelan2018, ChandraMisra2014, doi:10.1177/2158244020941844}. In addition, there are also concerns relating to privacy, internet access, interface design, counselor's skill, client's willingness, the platform of choice, and many more which need to be appropriately addressed for such intervention to have good effect\cite{10.1145/3516875.3516986, Sampson2020}.}
One of the participants also mentioned the possible use of AI/ML techniques for mitigating challenges in the counseling process. Multiple methods have been explored where AI and ML can be used for providing career counseling some of which include using a chatbot to conduct the Holland test (RIASEC test) and big 5 test for identifying personality followed by suggesting job opportunities \cite{10.1007/978-981-15-3242-9_1}, utilizing social media messaging applications \cite{10.1145/3484824.3484875}, to AI-powered career counseling for trainees graduating higher secondary school  \cite{10426016, 9441978} and many more \cite{9202024, Explaina31:online}. 
However, very little work has been done to identify the efficacy and impact of utilizing algorithmic systems for career counseling on vocational training especially for low-literate, low-income, and marginalized communities in the global south and India. In addition, though such technologies may prove useful, they must be dealt with caution as prior research suggests that blindly applying AI/ML technologies may cause more harm as most AI models are still black boxes and prone to failure. Prior work shows that persons with limited digital and low literacy might have incorrect perceptions of the fact-fulness of these models which may result in users blindly trusting these models often assuming that the machine’s expertise outweighed their own \cite{10.1145/3411764.3445420}. A study also demonstrates that it may not be fully possible to address bias issues like gender in AI recommendations without addressing the bias in humans \cite{10.1145/3490099.3511108}.
Despite these concerns, acknowledging the fact that the task of counseling at present is tedious, time-consuming, and detrimental to beneficiaries, it may be prudent to design and implement explainable, rule-based, non-probabilistic systems whose underlying processes are well-understood. Additionally, for probabilistic and black box systems, the trainees should have clear disclaimers on the potential errors, preferably through a human moderator/counselor/trainer to highlight its capabilities, limitations, and flaws \update{to give them a fair understanding of their options and make an informed decision. More importantly, irrespective of what kind of systems are designed they must be rigorously tested before deployment to avoid any potential harms.}

The learnings and findings from this study are not just limited to the Livelihood College at Dhamtari but are relevant to the rest of the Livelihood colleges in Chhattisgarh and shed light on the myriad of challenges that are faced while implementing Skill Development programs throughout the country. With multiple skill development schemes being launched at the state as well as National levels, it is pertinent to ensure that these programs are inclusive, high-quality, result-oriented, and one that can generate meaningful livelihoods for their beneficiaries. Even though this study sheds light on the challenges being faced in the first 3-steps of mobilization, counseling, and training stages in the Skill Development programs by the college authorities, it is still hard to ascertain what may be the actual reasons for some of the identified challenges such as the lack of trainees presence in the classrooms. Additionally, seeking information on the quality of training, and placement, and tracking from the college authorities may not provide an actual representation of reality as they may be inherently biased to respond positively. Hence, the entire picture can be further clarified from inputs from the beneficiaries of the Livelihood College who are in a better position to share a realistic view of the outcomes from training, placement, and tracking stages of the skill development programs. \update{Finally, many of the tech-oriented solutions discussed need further in-vivo testing before they can be deployed on scale and we encourage future empirical research including controlled trials, longitudinal, interventional studies, and other methods to explore causal relationships while adhering to ethical standards and test, refine/refute such ideas.}

\update{In summary, our findings suggest that skill development program implementation still faces a myriad of on-ground challenges that are not necessarily reflected at the policy-making or bureaucratic level highlighting inefficiencies, bottlenecks, and delays in information transmission. Drawing on amplification theory, we affirm that technology cannot be a substitute for institutional capacity and human intent \cite{10.1145/1940761.1940772}. While we did not find significant evidence that would bring the intent of the stakeholders we interacted with into question, the findings on lack of inclusive mobilization, a resource-constrained counseling process, sub-optimal use of digital assets, and the inadequate presence of trainees in classrooms, indicate missing institutional elements that can enable a seamless, efficient information exchange and prompt decision-making in current skill development programs. Technology here may be used as an enabler to support service delivery, however, establishing robust two-way communication channels among all stakeholders would go a long way in delivering better outcomes.}


\section{Conclusion}

This study looks into the service delivery of the Skill Development Program under Govt of Chhattisgrah’s Livelihood College. \update{By following the immersion/ crystallization approach,} we engaged in unstructured conversations and participant observations with administrators, trainers, mobilizers, and counselors of the Livelihood College of Dhamtari district along with nearby industry personnel, over one year. We also performed quantitative analysis and GIS mapping to further triangulate the qualitative inputs.
\update{Addressing \textbf{RQ1} on identifying challenges and bottlenecks in the implementation of the 5-stage process, first, we find a lack of inclusive mobilization, especially for rural inhabitants and gendered access to skill training arising from un-operational accommodation facilities and a lack of access to transportation services for trainees from far-flung regions of the district. Second, a tedious counseling process owing to its time-consuming and manual nature. Additionally, the lack of adequate support staff burdens the trainers with responsibilities that impact both the counseling as well as training activities. Third, we also find that there is inadequate trainee attendance/presence in the classrooms indicating that trainees may not be receiving a prescribed amount of training, thereby damaging their prospects. With regards to \textbf{RQ2} on how digital tools are implemented, we found sub-optimal implementation, and utilization of digital assets like biometric attendance, and CCTV installations. Following the findings, we discuss prospective methods to improve access to skill training} by empowering the Vocational Training Partners (VTPs) using the latest developments in digital technologies. We also discuss and recommend ways to improve the implementation and utilization of existing digital assets like biometric attendance and CCTV. Lastly, we discuss key considerations \update{while designing, deploying, and implementing solutions for improving the counseling process. We conclude by summarizing that skill development programs in India in particular the Livelihood College lack institution elements that allow for robust 2-way communication between stakeholders often resulting in inefficiencies and sub-standard service delivery. By addressing these issues, the program stands a higher chance of enhancing trainee engagement, ensuring more inclusive and effective skill training, and ultimately leading to better outcomes for the beneficiaries.}

\section{Limitations}
The authors acknowledge some limitations of this study. The current study looks into the challenges in implementing skill development programs from the perspectives of the administration, employees, and trainers of the Livelihood College. However, the views of an important stakeholder, i.e. the trainees who are the beneficiaries of the program were not represented in this study as their involvement in the implementation of the mobilization and counseling, and partly in the training process was minimal. Future work should look into understanding the perspectives of the trainees, their wants, needs, concerns, and constraints which is essential in delivering better outcomes. Additionally, our study only looks into the challenges faced at only one Livelihood College, and one may raise an argument that the findings from this work may not represent the rest of the livelihood college. On this point, we argue that the livelihood Colleges throughout the state of Chhattisgarh function under similar capacities and constraints, and the challenges identified are not particular to the district of Dhamtari. Our interactions with the college administrators and employees also confirmed the same. Having said this, we also acknowledge the limited sample size and it would surely have been beneficial if inputs could have been gathered from a set of Livelihood colleges from different districts allowing one to make a relative comparison of their performances. While our work takes the first step in this direction, we encourage more such empirical work taking into consideration different stakeholders across various districts to ensure that the intended beneficiaries of such programs are not left behind.


%
%
%
\begin{credits}
\subsubsection{\ackname} We thank our participants who graciously shared their insights, and experiences, and made this work possible. We also extend our thanks to the Mahatma Gandhi National Fellowship and the Office of District Skill Development Authority, Dhamtari, Chhattisgarh, India for supporting this work. Our heartfelt thanks to Professor Anirudha Joshi, Professor, IDC, IIT Bombay, for shepherding the paper and Professor Aditya Vashistha, Assistant Professor, Cornell University, for guiding us throughout the process. Finally, we thank the reviewers for their thoughtful comments and feedback on the paper. 
\end{credits}

\bibliographystyle{splncs04}
\bibliography{bibliography} 

@misc{Youthcan75,
  author =    {Meenakshi Datta Ghosh},
  title =  {Youth can be a clear advantage for {I}ndia- The Hindu},
  url =    {https://www.thehindu.com/opinion/lead/youth-can-be-a-clear-advantage-for-india/article30897179.ece},
  note = {last accessed on 11/03/2024}
  }

@misc{skilldev,
  key =    {Skill India},
  title =  {Skill {I}ndia, Strengthening New {I}ndia, GoI-MSDE},
  url =    {https://www.skilldevelopment.gov.in/sites/default/files/2020-01/English-ebook.pdf},
  note = {last accessed 11/03/2024}
  }

@misc{SkillInd58:online,
key = {Skill India, Economic Times},
title = {Skill {I}ndia, Digital Platform, Skill {I}ndia Digital platform launched to bring skilling initiatives under single umbrella, ET Government},
howpublished = {\href{https://government.economictimes.indiatimes.com/news/education/skill-india-digital-platform-launched-to-bring-skilling-initiatives-under-a-single-umbrella/103648139}{https://government.economictimes.indiatimes.com/news/education/skill-india-digital-platform-launched-to-bring-skilling-initiatives-under-a-single-umbrella/103648139}},
month = {},
year = {},
note = {last accessed 12/05/2024}
}

@misc{skilldevmission,
  author =    {GoI-MSDE},
  title =  {National Skill Development Mission},
  url =    {https://www.msde.gov.in/sites/default/files/2019-09/National\%20Skill\%20Development\%20Mission.pdf},
  note = {last accessed 11/03/2024}
  }

@misc{skillgapnsdc,
  author =    {NSDC},
  title =  {{D}istrict wise skill gap study for the state of {C}hhattisgarh},
  url =    {https://skillsip.nsdcindia.org/sites/default/files/kps-document/chattisgarh-district-skill-gap-study-final-report_18thJune.pdf},
  note = {last accessed 11/03/2024}
  }

@misc{HomePage84:online,
  key = {SPLCS},
  title =  {State Project Livelihood College Society-Home Page},
  url =    {https://splcs.cg.nic.in/},
  note = {last accessed 11/03/2024}
  }

@misc{Monitori50:online,
  author =    {MSDE},
  title =  {Monitoring Guidelines PMKVY 3.0},
  url =    {https://msde.gov.in/sites/default/files/2021-04/Monitoring%20Guidelines_PMKVY%203.0.pdf},
  note = {last accessed 11/03/2024}
  }

@misc{PMKVYGui27:online,
  author =    {MSDE},
  title =  {PMKVY Guideline report\_(06-01-2021)\_V5},
  url =    {https://www.msde.gov.in/sites/default/files/2021-01/PMKVY%20Guideline%20report_(06-01-2021)_V5.pdf},
  note = {last accessed 11/03/2024}
  }

@misc{SkillSaa82:online,
  key =    {Skill Saathi},
  title =  {Skill-Saathi-Guidelines},
  url =    {https://www.skilldevelopment.gov.in/sites/default/files/2019-12/Skill-Saathi-Guidelines.pdf},
  note = {last accessed 11/03/2024}
  }

@misc{Dhantari14:online,
  key =    {{D}hamtari, Government of {C}hhattisgarh, {I}ndia},
  title =  {{D}hamtari District | Land of Sacred Pond | {I}ndia},
  url =    {https://dhamtari.gov.in/},
  note = {last accessed 11/03/2024}
  }

@misc{IndiaCen49:online,
  key =    {India Census},
  title =  {India - Census of {I}ndia 2011 - {C}hhattisgarh - Series 23 - Part XII A - District Census Handbook, {D}hamtari},
  url =    {https://censusindia.gov.in/nada/index.php/catalog/334},
  note = {last accessed 11/03/2024}
  }

@misc{mospigov41:online,
author = {Goi-MOSPI},
title = {Periodic Labor Force Survey, NSSO, July 2022 - June 2023},
howpublished = {\url{https://mospi.gov.in/sites/default/files/publication_reports/AR_PLFS_2022_23N.pdf?download=1}},
month = {},
year = {},
note = {last accessed 12/05/2024}
}

@misc{BlockPan6:online,
key = {Block and Panchayat, Dhamtari},
title = {Block \& Panchayat | {D}hamtari District | {I}ndia},
howpublished = {\url{https://dhamtari.gov.in/en/about-district/administrative-setup/block/}},
month = {},
year = {},
note = {last accessed 14/05/2024}
}

@misc{Unemploy59:online,
Key = {Unemployment Rate, Forbes {I}ndia},
title = {Unemployment Rate In {I}ndia(2008 To 2024): Current Rate, Historical Trends And More - Forbes {I}ndia},
howpublished = {\url{https://www.forbesindia.com/article/explainers/unemployment-rate-in-india/87441/1}},
month = {},
year = {},
note = {last accessed 12/05/2024}
}

@misc{Populati31:online,
key = {Population ages - World Bank},
title = {Population ages 15-64 (\% of total population) - {I}ndia| Data},
howpublished={\url{https://data.worldbank.org/indicator/SP.POP.1564.TO.ZS?locations=IN}},
month = {},
year = {},
note = {last accessed 12/05/2024}
}

@article{HOSAN2022129858,
title = {Dynamic links among the demographic dividend, digitalization, energy intensity and sustainable economic growth: Empirical evidence from emerging economies},
journal = {Journal of Cleaner Production},
volume = {330},
pages = {129858},
year = {2022},
issn = {0959-6526},
doi = {https://doi.org/10.1016/j.jclepro.2021.129858},
url = {https://www.sciencedirect.com/science/article/pii/S0959652621040300},
author = {Shahadat Hosan and Shamal Chandra Karmaker and Md Matiar Rahman and Andrew J. Chapman and Bidyut Baran Saha}
}

@Article{Demograp14:online,
author={Jafrin, Nusrat
and Mahi, Masnun
and Masud, Muhammad Mehedi
and Ghosh, Deboshree},
title={Demographic dividend and economic growth in emerging economies: fresh evidence from the SAARC countries},
journal={International Journal of Social Economics},
year={2021},
month={Jan},
day={01},
publisher={Emerald Publishing Limited},
volume={48},
number={8},
pages={1159-1174},
issn={0306-8293},
doi={10.1108/IJSE-08-2020-0588},
url={https://doi.org/10.1108/IJSE-08-2020-0588}
}

@article{BukuPoli36:online,
  title={Taking advantage of the demographic dividend in Indonesia},
  author={Hayes, Adrian and Setyonaluri, Diahhadi},
  journal={A Brief Introduction to Theory and Practice},
  year={2015},
  publisher={The United Nations Population Fund Jakarta}
}

@misc{chhattis96:online,
key = {Chhattisgarh Skill Development Act, India Code},
title = {Chhattisgarh Right of youth to skill development act, 2013 no. 17 of 2013 date 26.04.2013},
howpublished = {\url{https://www.indiacode.nic.in/bitstream/123456789/12618/1/chhattisgarh_right_of_youth_to_skill_development_act%2c_2013_no._17_of_2013_date_26.04.2013.pdf}},
month = {},
year = {},
note = {last accessed 12/05/2024}
}

@inproceedings{10.1145/3332165.3347889,
author = {Ahuja, Karan and Harrison, Chris and Goel, Mayank and Xiao, Robert},
title = {MeCap: Whole-Body Digitization for Low-Cost VR/AR Headsets},
year = {2019},
isbn = {9781450368162},
publisher = {Association for Computing Machinery},
address = {New York, NY, USA},
url = {https://doi.org/10.1145/3332165.3347889},
doi = {10.1145/3332165.3347889},
booktitle = {Proceedings of the 32nd Annual ACM Symposium on User Interface Software and Technology},
pages = {453–462},
numpages = {10},
location = {New Orleans, LA, USA},
series = {UIST '19}
}

@article{https://doi.org/10.1111/ajae.12114,
author = {Anukriti, S and Herrera-Almanza, Catalina and Pathak, Praveen K. and Karra, Mahesh},
title = {Curse of the Mummy-ji: The Influence of Mothers-in-Law on Women in {I}ndia†},
journal = {American Journal of Agricultural Economics},
volume = {102},
number = {5},
pages = {1328-1351},
doi = {https://doi.org/10.1111/ajae.12114},
eprint = {https://onlinelibrary.wiley.com/doi/pdf/10.1111/ajae.12114},
year = {2020}
}

@article{BALACHANDRAN2024100060,
title = {Transportation, employment and gender norms: Evidence from {I}ndian cities},
journal = {Regional Science Policy \& Practice},
pages = {100060},
year = {2024},
issn = {1757-7802},
doi = {https://doi.org/10.1016/j.rspp.2024.100060},
author = {Arun Balachandran and Sonalde Desai}
}

@inproceedings{10.1145/3313831.3376674,
author = {Thakkar, Divy and Kumar, Neha and Sambasivan, Nithya},
title = {Towards an AI-powered Future that Works for Vocational Workers},
year = {2020},
isbn = {9781450367080},
publisher = {Association for Computing Machinery},
address = {New York, NY, USA},
url = {https://doi.org/10.1145/3313831.3376674},
doi = {10.1145/3313831.3376674},
booktitle = {Proceedings of the 2020 CHI Conference on Human Factors in Computing Systems},
pages = {1–13},
numpages = {13},
location = {<conf-loc>, <city>Honolulu</city>, <state>HI</state>, <country>USA</country>, </conf-loc>},
series = {CHI '20}
}

@article{CHIANG2022107125,
title = {Augmented reality in vocational training: A systematic review of research and applications},
journal = {Computers in Human Behavior},
volume = {129},
pages = {107125},
year = {2022},
issn = {0747-5632},
doi = {https://doi.org/10.1016/j.chb.2021.107125},
author = {Feng-Kuang Chiang and Xiaojing Shang and Lu Qiao}
}

@InProceedings{10.1007/978-981-15-3242-9_1,
author="D'Silva, Godson
and Jani, Megh
and Jadhav, Vipul
and Bhoir, Amit
and Amin, Prithvi",
editor="Vasudevan, Hari
and Michalas, Antonis
and Shekokar, Narendra
and Narvekar, Meera",
title="Career Counselling Chatbot Using Cognitive Science and Artificial Intelligence",
booktitle="Advanced Computing Technologies and Applications",
year="2020",
publisher="Springer Singapore",
address="Singapore",
pages="1--9",
isbn="978-981-15-3242-9"
}

@INPROCEEDINGS{10426016,
  author={Ghuge, Madhuri and Kamble, Torana and Mandrawliya, Anushka and Kumari, Anupam and Raikwar, Vinay},
  booktitle={2023 3rd International Conference on Innovative Mechanisms for Industry Applications (ICIMIA)}, 
  title={Envisioning Tomorrow: AI Powered Career Counseling}, 
  year={2023},
  volume={},
  number={},
  pages={377-383},
  doi={10.1109/ICIMIA60377.2023.10426016}}

@inproceedings{10.1145/3484824.3484875,
author = {Suresh, Nalina and Mukabe, Nkandu and Hashiyana, Valerianus and Limbo, Anton and Hauwanga, Aina},
title = {Career Counseling Chatbot on Facebook Messenger using AI},
year = {2022},
isbn = {9781450387637},
publisher = {Association for Computing Machinery},
address = {New York, NY, USA},
url = {https://doi.org/10.1145/3484824.3484875},
doi = {10.1145/3484824.3484875},
booktitle = {Proceedings of the International Conference on Data Science, Machine Learning and Artificial Intelligence},
pages = {65–73},
numpages = {9},
location = {<conf-loc>, <city>Windhoek</city>, <country>Namibia</country>, </conf-loc>},
series = {DSMLAI '21'}
}

@INPROCEEDINGS{9441978,
  author={Vignesh, S and Shivani Priyanka, C and Shree Manju, H and Mythili, K},
  booktitle={2021 7th International Conference on Advanced Computing and Communication Systems (ICACCS)}, 
  title={An Intelligent Career Guidance System using Machine Learning}, 
  year={2021},
  volume={1},
  number={},
  pages={987-990},
  doi={10.1109/ICACCS51430.2021.9441978}}

@INPROCEEDINGS{9202024,
  author={Joshi, Kartikey and Goel, Amit Kumar and Kumar, Tapas},
  booktitle={2020 7th International Conference on Smart Structures and Systems (ICSSS)}, 
  title={Online Career Counsellor System based on Artificial Intelligence: An approach}, 
  year={2020},
  volume={},
  number={},
  pages={1-4},
  doi={10.1109/ICSSS49621.2020.9202024}}

@article{Explaina31:online,
author = {Guleria, Pratiyush and Sood, Manu },
journal= {Education and Information Technologies},
title = {Explainable AI and machine learning: performance evaluation and explainability of classifiers on educational data mining inspired career counseling | Education and Information Technologies},
doi={https://doi.org/10.1007/s10639-022-11221-2},
month = {},
year = {2023},
volume = {28},
pages = {1081–1116}}

@inproceedings{10.1145/3411764.3445420,
author = {Okolo, Chinasa T. and Kamath, Srujana and Dell, Nicola and Vashistha, Aditya},
title = {“It cannot do all of my work”: Community Health Worker Perceptions of AI-Enabled Mobile Health Applications in Rural {I}ndia},
year = {2021},
isbn = {9781450380966},
publisher = {Association for Computing Machinery},
address = {New York, NY, USA},
url = {https://doi.org/10.1145/3411764.3445420},
doi = {10.1145/3411764.3445420},
booktitle = {Proceedings of the 2021 CHI Conference on Human Factors in Computing Systems},
articleno = {701},
numpages = {20},
location = {<conf-loc>, <city>Yokohama</city>, <country>Japan</country>, </conf-loc>},
series = {CHI '21}
}

@Article{RePEc:fas:journl:v:13:y:2023:i:2:p:4-28,
  author={Arindam Das and Yoshifumi Usami},
  title={{Downturn in Wages in Rural {I}ndia}},
  journal={Journal},
  year=2023,
  volume={13},
  number={2},
  pages={4-28},
  month={July-Dece},
  doi={},
  url={https://ideas.repec.org/a/fas/journl/v13y2023i2p4-28.html}
}

@article{Kunal2022EmployerAG,
  title={Employer attractiveness: generation z employment expectations in {I}ndia},
  author={Kumar Kunal and Philip Coelho and Somu Pooja},
  journal={CARDIOMETRY},
  year={2022},
  url={https://api.semanticscholar.org/CorpusID:252243610}
}

@article{doi:10.1177/18479790221112548,
author = {Thang Nguyen Ngoc and Mai Viet Dung and Chris Rowley and Mirjana Pejić Bach},
title ={Generation Z job seekers’ expectations and their job pursuit intention: Evidence from transition and emerging economy},

journal = {International Journal of Engineering Business Management},
volume = {14},
number = {},
pages = {18479790221112548},
year = {2022},
doi = {10.1177/18479790221112548},

URL = { 
    
        https://doi.org/10.1177/18479790221112548
    
    

},
eprint = { 
    
        https://doi.org/10.1177/18479790221112548
    
    

}
}

@INPROCEEDINGS{7808130,
  author={Veer, Nilesh D. and Momin, B. F.},
  booktitle={2016 IEEE International Conference on Recent Trends in Electronics, Information \& Communication Technology (RTEICT)}, 
  title={An automated attendance system using video surveillance camera}, 
  year={2016},
  volume={},
  number={},
  pages={1731-1735},
  doi={10.1109/RTEICT.2016.7808130}}

@ARTICLE{9610134,
  author={Su, Mu-Chun and Cheng, Chun-Ting and Chang, Ming-Ching and Hsieh, Yi-Zeng},
  journal={IEEE Transactions on Consumer Electronics}, 
  title={A Video Analytic In-Class Student Concentration Monitoring System}, 
  year={2021},
  volume={67},
  number={4},
  pages={294-304},
  doi={10.1109/TCE.2021.3126877}}

@INPROCEEDINGS{9057967,
  author={Krishnnan, Nirmal and Ahmed, Saeed and Ganta, Thanmay and Jeyakumar, Gurusamy},
  booktitle={2020 10th International Conference on Cloud Computing, Data Science \& Engineering (Confluence)}, 
  title={A Video Analytics Based Solution for Detecting the Attention Level of the Students in Class Rooms}, 
  year={2020},
  volume={},
  number={},
  pages={498-501},
  doi={10.1109/Confluence47617.2020.9057967}}

@ARTICLE{Wu2023-mw,
  title    = "A privacy-preserving student status monitoring system",
  author   = "Wu, Haopeng and Lu, Zhiying and Zhang, Jianfeng",
  journal  = "Complex \& Intelligent Systems",
  volume   =  9,
  number   =  1,
  pages    = "597--608",
  month    =  feb,
  year     =  2023
}

@ARTICLE{Preuveneers2021-ud,
  title    = "Cloud and edge based data analytics for privacy-preserving
              multi-modal engagement monitoring in the classroom",
  author   = "Preuveneers, Davy and Garofalo, Giuseppe and Joosen, Wouter",
  journal  = "Information Systems Frontiers",
  volume   =  23,
  number   =  1,
  pages    = "151--164",
  month    =  feb,
  year     =  2021
}

@inproceedings{10.1145/3490099.3511108,
author = {Wang, Clarice and Wang, Kathryn and Bian, Andrew and Islam, Rashidul and Keya, Kamrun Naher and Foulds, James and Pan, Shimei},
title = {Do Humans Prefer Debiased AI Algorithms? A Case Study in Career Recommendation},
year = {2022},
isbn = {9781450391443},
publisher = {Association for Computing Machinery},
address = {New York, NY, USA},
url = {https://doi.org/10.1145/3490099.3511108},
doi = {10.1145/3490099.3511108},
booktitle = {Proceedings of the 27th International Conference on Intelligent User Interfaces},
pages = {134–147},
numpages = {14},
location = {<conf-loc>, <city>Helsinki</city>, <country>Finland</country>, </conf-loc>},
series = {IUI '22}
}

@misc{Introduc1:online,
key = {Digital {I}ndia},
title = {Digital {I}ndia: {I}ntroduction},
howpublished = {\url{https://www.digitalindia.gov.in/introduction/}},
month = {},
year = {},
note = {last accessed 26/05/2024}
}

@misc{DigitalI21:online,
key = {Digital India Week},
title = {Digital {I}ndia Week: Digital Locker, MyGov.in, and other projects that were unveiled | Technology News - The {I}ndian Express},
howpublished = {\url{https://indianexpress.com/article/technology/tech-news-technology/projects-and-policies-launched-at-digital-india-week/}},
month = {},
year = {},
note = {last accessed 27/05/2024}
}

@article{ghosh2022skilling,
  title={Skilling the {I}ndian youth: a State-level analysis},
  author={Ghosh, Piyali and Goel, Geetika and Bhongade, Ankita},
  journal={Benchmarking: An International Journal},
  volume={29},
  number={10},
  pages={3379--3395},
  year={2022},
  publisher={Emerald Publishing Limited}
}

@article{agrawal2020demographic,
  title={Demographic Dividend: Skill Development Evidence in {I}ndia},
  author={Agrawal, Mini and Singh, Chetanya and Thakur, KS},
  journal={Available at SSRN 3590719},
  year={2020}
}

@article{Kumar_Nain_Singh_Kumbhare_Parsad_Kumar_2021,
  author = {Kumar, G.S.A. and Nain, M.S. and Singh, R. and Kumbhare, N.V. and Parsad, R. and Kumar, S.},
  title = {Training Effectiveness of Skill Development Training Programmes among the Aspirational Districts of Karnataka},
  journal = {Indian Journal of Extension Education},
  volume = {57},
  number = {4},
  pages = {67--70},
  year = {2021},
  translator = {Kumar, S.},
  doi = {10.48165/IJEE.2021.57415},
  url = {https://doi.org/10.48165/IJEE.2021.57415}
}

@article{sinha2022impact,
  title={Impact Analysis of Skill Development on the Performance of Small Tea Growers of Assam},
  author={Sinha, Sanjay},
  journal={Pacific Business Review (International)},
  volume={14},
  number={7},
  pages={97--108},
  year={2022}
}

@article{chakravorty2019skills,
  title={Skills training and employment outcomes in rural Bihar},
  author={Chakravorty, Bhaskar and Bedi, Arjun S},
  journal={The {I}ndian Journal of Labour Economics},
  volume={62},
  pages={173--199},
  year={2019},
  publisher={Springer}
}

@misc{NEPFinal2:online,
author = {MHRD, Government of {I}ndia},
title = {New {E}ducation {P}olicy},
howpublished = {\url{https://www.education.gov.in/sites/upload_files/mhrd/files/NEP_Final_English.pdf}},
month = {},
year = {2020},
note = {last accessed 01/06/2024}
}

@article{Waoo_Waoo_2022, 
    title={The New Educational Policy in {I}ndia: Towards a Digital Future}, 
    volume={2}, 
    DOI={10.37899/journallaedusci.v2i6.558}, 
    abstractNote={&lt;p&gt;&lt;em&gt;New Education Policy was formulated by the Government via a consultation process. It emerges as an inclusive, participatory, and holistic approach of MHRD initiated in January 2015, MHRDA new National Education Policy (NEP) has been approved by the Union Cabinet, which makes modifications in the {I}ndian system of education from school to college. The New Education Policy outlines India’s goal of becoming a knowledge superpower. As part of the restructuring, the Ministry of Human Resources Development (MHRD) became the Ministry of Education. NEP promotes ideas, concepts, applications, and problem-solving activities. This policy calls for more interactive teaching-learning. A technology-based educational approach is emphasized in this policy. This policy demonstrates a greater use of ICT for remote and interactive education at school and in higher education. The study in this paper shows the impact of ICT tools on future education and various methods for building virtual infrastructure for learning.&lt;/em&gt;&lt;/p&#38;gt;}, 
    number={6}, 
    journal={Journal La Edusci}, 
    author={Waoo, Dr. Akhilesh A. and Waoo, Dr. Ashwini A.}, year={2022}, 
    month={Mar.}, 
    pages={30-34}
}

@techreport{Swain2009impact,
address = {Uppsala},
author = {Ranjula Bali Swain and Adel Varghese},
copyright = {http://www.econstor.eu/dspace/Nutzungsbedingungen},
language = {eng},
note = {urn:nbn:se:uu:diva-106712},
number = {2009:11},
publisher = {Uppsala University, Department of Economics},
title = {The impact of skill development and human capital training on self help groups},
type = {Working Paper},
url = {https://hdl.handle.net/10419/82586},
institution = {Uppsala University, Department of Economics},
year = {2009}
}

@article{roy2018knowledge,
  title={Knowledge and skill development of Bihar farmers on inland fisheries management: A terminal evaluation},
  author={Roy, Aparna and Das, BK and Chandra, Ganesh and Das, Archan Kanti and Raman, RK},
  journal={Indian Journal of Fisheries},
  volume={65},
  number={2},
  pages={119--123},
  year={2018}
}

@article{agrawal2019impact,
  title={Impact of Pradhan Mantri Kaushal Vikas Yojana on the Productivity of Youth in Gwalior Region, {I}ndia},
  author={Agrawal, Mini and Thakur, KS},
  journal={International Journal of Recent Technology and Engineering (IJRTE), 8 (4), 801},
  volume={806},
  year={2019}
}

@article{Tiwari_Malati_2020, title={Employability Skill Evaluation Among Vocational Education Students in {I}ndia}, volume={12},  url={https://publisher.uthm.edu.my/ojs/index.php/JTET/article/view/4873}, abstractNote={The changing nature of work and employment is providing individuals for more flexible multi-skilling and learning opportunities. Imparting skill-based industry-oriented teaching can bridge the skill gaps and enhance employment opportunities for students. In this context, Government of {I}ndiahas introduced numerous programs to provide a fillip to technical vocational education and training.&amp;amp;nbsp; The current paper is aimed at understanding the role of vocational education and the change it brings to skill development and employability of the students.&amp;amp;nbsp; A combination of both qualitative and quantitative research methods was deployed for the study.&amp;amp;nbsp; In the quantitative design a multi-stage sampling process comprising of both probabilistic and non-probabilistic methods was employed.&amp;amp;nbsp; A sample of 586 students pursuing retail vocational education was identified and administered with the questionnaire.&amp;amp;nbsp; The statistical analysis presented the socio-economic profiles.&amp;amp;nbsp; Further, five factors for skill development and one factor for employability skill were identified through exploratory factor analysis.&amp;amp;nbsp; The factors identified for skill development include Initiative and Enterprise Skills (IES), Workplace Skills (WS), Professional Practice and Standards (PPS), Inter Personal Skills (IPS) and Integration Theory and Practice (ITP).&amp;amp;nbsp; Confirmatory and regression models involving all the factors were tested and their significance was analyzed.&amp;amp;nbsp; The study revealed that there is a positive impact of skill development on employability.&amp;amp;nbsp; It is suggested that focus on imparting vocational education for skill development can be a panacea for increased employability.&amp;amp;nbsp;&amp;amp;nbsp;}, number={1}, journal={Journal of Technical Education and Training}, author={Tiwari, Pratiksha and Malati, Nittala}, year={2020}, month={Jan.} }

@article{Pandey_Gupta_Pandey_Singh_2017,
  author = {Pandey, A. and Gupta, N. and Pandey, A. and Singh, S.},
  title = {Impact of Vocational Training on Value Addition in Knowledge and Adoption of Rural Women},
  journal = {Indian Journal of Extension Education},
  volume = {53},
  number = {3},
  pages = {36--39},
  year = {2017},
  url = {https://acspublisher.com/journals/index.php/ijee/article/view/4977}
}

@misc{MapofDis47:online,
key = {Map of Dhamtari, Government of {C}hhattisgarh},
title = {Map of District | {D}hamtari District | {I}ndia},
howpublished = {\url{https://dhamtari.gov.in/en/about-district/map-of-district/}},
month = {},
year = {},
note = {last accessed 01/06/2024}
}

@misc{YOLOv5Ul97:online,
key = {Yolov5},
title = {YOLOv5 - Ultralytics YOLO Docs},
howpublished = {\url{https://docs.ultralytics.com/models/yolov5/#performance-metrics}},
month = {},
year = {},
note = {last accessed 01/06/2024}
}

@misc{WhatisAa54:online,
author = {UIDAI},
title = {What is Aadhaar? - Unique Identification Authority of {I}ndia| Government of {I}ndia},
howpublished = {\url{https://uidai.gov.in/en/16-english-uk/aapka-aadhaar/14-what-is-aadhaar.html}},
month = {},
year = {},
note = {last accessed 09/06/2024}
}

@misc{PMKVYpdf14:online,
author = {National Skill Development Corporation, Ministry of Skill Development \& Entrepreneurship, 
Government of {I}ndia},
title = {PMKVY},
howpublished = {\url{https://www.mofpi.gov.in/sites/default/files/PMKVY.pdf}},
month = {},
year = {2015},
note = {last accessed 10/06/2024}
}

@inproceedings{10.1145/3596671.3597655,
author = {Kim, Pyeonghwa and Sawyer, Steve},
title = {Many Futures of Work and Skill: Heterogeneity in Skill Building Experiences on Digital Labor Platforms},
year = {2023},
isbn = {9798400708077},
publisher = {Association for Computing Machinery},
address = {New York, NY, USA},
url = {https://doi.org/10.1145/3596671.3597655},
doi = {10.1145/3596671.3597655},
booktitle = {Proceedings of the 2nd Annual Meeting of the Symposium on Human-Computer Interaction for Work},
articleno = {11},
numpages = {9},
location = {<conf-loc>, <city>Oldenburg</city>, <country>Germany</country>, </conf-loc>},
series = {CHIWORK '23}
}

@article{Sheshadri_Pradeep_Chandran_2021, title={Towards Gender Inclusive Skill Development in Rural {I}ndia: Factors that Inhibit and Facilitate Skill Womenâ€™s Enrolment in Vocational Training}, volume={6}, url={https://ebpj.e-iph.co.uk/index.php/EBProceedings/article/view/3032}, DOI={10.21834/ebpj.v6iSI4.3032}, abstractNote={&amp;lt;p&amp;gt;&amp;lt;a style=&amp;quot;cursor: pointer;&amp;quot; href=&amp;quot;https://crossmark.crossref.org/dialog?doi=10.21834/ebpj.v6iSI4.3032&amp;amp;amp;domain=ebpj.e-iph.co.uk&amp;amp;amp;uri_scheme=http%3A&amp;amp;amp;cm_version=v2.0&amp;quot; data-target=&amp;quot;crossmark&amp;quot;&amp;gt;&amp;lt;img src=&amp;quot;https://ebpj.e-iph.co.uk/xmark.jpg&amp;quot; alt=&amp;quot;&amp;quot; /&amp;gt; &amp;lt;/a&amp;gt;&amp;lt;/p&amp;gt;
&amp;lt;p&amp;gt;Women in rural {I}ndiaare the country’s most underserved population regarding access to skill development opportunities. Despite rhetoric at the national and international policy levels acknowledging the dearth of female participation in vocational training and subsequently skilled labour in India, female enrolment remains low. A greater understanding of factors that facilitate and hinder women’s enrolment in skill development programs, particularly in the current era of pro-skill development, where vocational training is highly subsidized, if not free of cost, is required to design effective interventions that are inclusive of this perpetually side-lined population. Towards developing this improved understanding, an exploratory qualitative study was conducted in the rural {I}ndian village of Juna Khatiwada, Madhya Pradesh, where vocational training programs for women have been made available and accessible, free of cost. Semi-structured interviews and focused group discussions were conducted with 16 women of Juna Khatiwada. In addition to identifying factors that facilitate and inhibit enrolment among the target population, the study also revealed that women who did enrol and complete vocational training courses reported better coping with domestic economic challenges. Findings from this study serve to provide recommendations on the way forward in terms of skill development policy and practice that are more inclusive of women in rural {I}ndian.&amp;lt;/p&amp;gt;
&amp;lt;p&amp;gt;Keywords: Education and Training; Employment; Technical Vocational Rural {I}ndia; Women&amp;lt;/p&amp;gt;
&amp;lt;p&amp;gt;&amp;lt;em&amp;gt;eISSN: 2398-4287Â© 2021. The Authors. Published for AMER ABRA cE-Bs by e-International Publishing House, Ltd., UK. This is an open access article under the CC BYNC-ND license (http://creativecommons.org/licenses/by-nc-nd/4.0/). Peerâ€“review under responsibility of AMER (Association of Malaysian Environment-Behaviour Researchers), ABRA (Association of Behavioural Researchers on Asians/Africans/Arabians) and cE-Bs (Centre for Environment-Behaviour Studies), Faculty of Architecture, Planning &amp;amp;amp; Surveying, Universiti Teknologi MARA, Malaysia.&amp;lt;/em&amp;gt;&amp;lt;/p&amp;gt;
&amp;lt;p&amp;gt;DOI: https://doi.org/10.21834/ebpj.v6iSI4.3032&amp;lt;/p&amp;gt;}, number={SI4}, journal={Environment-Behaviour Proceedings Journal}, author={Sheshadri, Srividya and Pradeep, Ayswarya and Chandran , Mamatha}, year={2021}, month={Jul.}, pages={239–243} }

@article{doi:10.1080/01436597.2022.2077184,
author = {Carol Upadhya and Supriya RoyChowdhury},
title = {Crafting new service workers: skill training, migration and employment in Bengaluru, {I}ndia},
journal = {Third World Quarterly},
volume = {45},
number = {4},
pages = {753--770},
year = {2024},
publisher = {Routledge},
doi = {10.1080/01436597.2022.2077184},


URL = { 
    
        https://doi.org/10.1080/01436597.2022.2077184
    
    

},
eprint = { 
    
        https://doi.org/10.1080/01436597.2022.2077184
    
    

}

}

@inproceedings{10.1145/3380867.3426199,
author = {Agrawal, Richa and Pillai, Jayesh S.},
title = {Augmented Reality Application in Vocational Education: A Case of Welding Training},
year = {2020},
isbn = {9781450375269},
publisher = {Association for Computing Machinery},
address = {New York, NY, USA},
url = {https://doi.org/10.1145/3380867.3426199},
doi = {10.1145/3380867.3426199},
booktitle = {Companion Proceedings of the 2020 Conference on Interactive Surfaces and Spaces},
pages = {23–27},
numpages = {5},
location = {Virtual Event, Portugal},
series = {ISS '20}
}

@InProceedings{10.1007/978-3-319-67684-5_1,
author="Sachith, K. P.
and Gopal, Aiswarya
and Muir, Alexander
and Bhavani, Rao R.",
editor="Bernhaupt, Regina
and Dalvi, Girish
and Joshi, Anirudha
and K. Balkrishan, Devanuj
and O'Neill, Jacki
and Winckler, Marco",
title="Contextualizing ICT Based Vocational Education for Rural Communities: Addressing Ethnographic Issues and Assessing Design Principles",
booktitle="Human-Computer Interaction - INTERACT 2017",
year="2017",
publisher="Springer International Publishing",
address="Cham",
pages="3--12",
isbn="978-3-319-67684-5"
}

@INPROCEEDINGS{6700139,
  author={Akshay, Nagarajan and Deepu, Sasi and Rahul, E.S. and Ranjith, R. and Jose, James and Unnikrishnan, R. and Bhavani, Rao R.},
  booktitle={IECON 2013 - 39th Annual Conference of the IEEE Industrial Electronics Society}, 
  title={Design and evaluation of a Haptic simulator for vocational skill Training and Assessment}, 
  year={2013},
  volume={},
  number={},
  pages={6108-6113},
  doi={10.1109/IECON.2013.6700139}}

@INPROCEEDINGS{6208644,
  author={Akshay, Nagarajan and Sreeram, Kongeseri and Anand, Aneesh and Venkataraman, Ranga and Bhavani, Rao R.},
  booktitle={2012 IEEE International Conference on Technology Enhanced Education (ICTEE)}, 
  title={MoVE: Mobile vocational education for rural {I}ndia}, 
  year={2012},
  volume={},
  number={},
  pages={1-5},
  doi={10.1109/ICTEE.2012.6208644}}

@article{gig_economy, 
    author = {Biswabhushan Behera and Mamta Gaur}, 
    volume={13}, 
    title = {Skill Training for the Success of the Gig Economy},  
    DOI={10.47750/pnr.2022.13.S05.429}, 
    abstractNote={The world is currently witnessing a boom in the gig economy due to frequent market disruptions, technological advancement, and the aspirations of the millennials. There is a win-win situation for both employers and gig workers. Employers are excited about the reduced cost of employment and gig workers are excited due to increased freedom, flexibility, and finances. However, on the flip side is the lack of responsibility to manage the skill level with the technological advancement, and the future requirements of the market. The growing skill mismatch is obvious and calls for skill development of the gig workers for the success of the gig economy. The study has made an effort to understand the gig economy in India, its salient features and has deliberated on various aspects of the skill development for the gig workers like skill requirements, challenges involved and various strategies of skill development. The study has concluded that Gig workers need to understand their own responsibility towards upskilling themselves and keep learning to remain relevant. Employers should also understand their responsibility in skilling gig workers and familiarize them with the organization’s goals, strategy, culture, and operations. At the same time, they need to train their regular employees on collaborating with the gig workers so that both these groups may operate cohesively while acknowledging and respecting one another&amp;#039;s unique work ethos. The government needs to bring in the required policy framework to create a conducive environment for the gig economy. The success of the gig economy depends on the cohesive and collaborative efforts of both these groups of employees, employers and the Government.}, 
    journal={Journal of Pharmaceutical Negative Results}, 
    year={2022}, 
    month={Dec.}, 
    pages={2835–2840} 
}

@article{doi:10.1177/26314541211064726,
author = {Jeeva Balakrishnan},
title ={Building Capabilities for Future of Work in the Gig Economy},

journal = {NHRD Network Journal},
volume = {15},
number = {1},
pages = {56-70},
year = {2022},
doi = {10.1177/26314541211064726},

URL = { 
    
        https://doi.org/10.1177/26314541211064726
    
    

},
eprint = { 
    
        https://doi.org/10.1177/26314541211064726
    
    

}
}

@InProceedings{10.1007/978-3-030-90238-4_2,
author="Dugar, Sumesh
and Mitra, Abhishek
and Nandi, Shweta
and Adhikary, Biswajit
and Paul, Sonit
and Manusuriya, Madhav",
editor="Stephanidis, Constantine
and Soares, Marcelo M.
and Rosenzweig, Elizabeth
and Marcus, Aaron
and Yamamoto, Sakae
and Mori, Hirohiko
and Rau, Pei-Luen Patrick
and Meiselwitz, Gabriele
and Fang, Xiaowen
and Moallem, Abbas",
title="Interaction Design for the Next Billion Users",
booktitle="HCI International 2021 - Late Breaking Papers: Design and User Experience",
year="2021",
publisher="Springer International Publishing",
address="Cham",
pages="16--23",
isbn="978-3-030-90238-4"
}

@book{crabtree2023doing,
  title={Doing qualitative research},
  author={Crabtree, Benjamin F and Miller, William L},
  year={2023},
  publisher={Sage publications}
}

@misc{Globalla24:online,
key = {Gloabl Labour market - International Labour Organization},
title = {Global labour market to deteriorate further as Ukraine conflict and other crises continue | International Labour Organization},
howpublished = {\href{https://www.ilo.org/resource/news/global-labour-market-deteriorate-further-ukraine-conflict-and-other-crises-0}{https://www.ilo.org/resource/news/global-labour-market-deteriorate-further-ukraine-conflict-and-other-crises-0}},
month = {},
year = {},
note = {last accessed 26/08/2024}
}

@misc{GlobalEc91:online,
author = {Pierre-Olivier Gourinchas},
title = {Global Economy on Track but Not Yet Out of the Woods},
howpublished = {\url{https://www.imf.org/en/Blogs/Articles/2023/07/25/global-economy-on-track-but-not-yet-out-of-the-woods}},
month = {},
year = {},
note = {last accessed 26/08/2024}
}

@misc{Technolo82:online,
author = {Devansh Pathak},
title = {Technology could disrupt the blue-collar workforce worldwide | World Economic Forum},
howpublished = {\href{https://www.weforum.org/agenda/2023/04/growth-summit-2023-technology-is-about-to-disrupt-the-blue-collar-workforce-in-emerging-markets/}{https://www.weforum.org/agenda/2023/04/growth-summit-2023-technology-is-about-to-disrupt-the-blue-collar-workforce-in-emerging-markets/}},
month = {April},
year = {2023},
note = {last accessed 26/08/2024}
}

@article{doi:10.1080/0145935X.2024.2340551,
author = {P. Behera, M. Singh and S. Tarai},
title = {Short-Term Vocational Courses as a Career-Building Program for the Youth in {C}hhattisgarh, {I}ndia},
journal = {Child \& Youth Services},
volume = {0},
number = {0},
pages = {1--24},
year = {2024},
publisher = {Routledge},
doi = {10.1080/0145935X.2024.2340551},


URL = { 
    
        https://doi.org/10.1080/0145935X.2024.2340551
    
    

},
eprint = { 
    
        https://doi.org/10.1080/0145935X.2024.2340551
    
    

}

}

@Article{Pordelan2018,
author={Pordelan, Nooshin
and Sadeghi, Ahmad
and Abedi, Mohammad Reza
and Kaedi, Marjan},
title={How online career counseling changes career development: A life design paradigm},
journal={Education and Information Technologies},
year={2018},
month={Nov},
day={01},
volume={23},
number={6},
pages={2655-2672},
issn={1573-7608},
doi={10.1007/s10639-018-9735-1},
url={https://doi.org/10.1007/s10639-018-9735-1}
}

@Article{ChandraMisra2014,
author={{Amritesh}
and Chandra Misra, Subhas
and Chatterjee, Jayanta},
title={Emerging scenario of online counseling services in {I}ndia: a case of e-government intervention},
journal={Transforming Government: People, Process and Policy},
year={2014},
month={Jan},
day={01},
publisher={Emerald Group Publishing Limited},
volume={8},
number={4},
pages={569-596},
issn={1750-6166},
doi={10.1108/TG-10-2013-0040},
url={https://doi.org/10.1108/TG-10-2013-0040}
}

@Article{Sampson2020,
author={Sampson, James P.
and Kettunen, Jaana
and Vuorinen, Raimo},
title={The role of practitioners in helping persons make effective use of information and communication technology in career interventions},
journal={International Journal for Educational and Vocational Guidance},
year={2020},
month={Apr},
day={01},
volume={20},
number={1},
pages={191-208},
issn={1573-1782},
doi={10.1007/s10775-019-09399-y},
url={https://doi.org/10.1007/s10775-019-09399-y}
}

@inproceedings{10.1145/3516875.3516986,
author = {Purwaningrum, Ribut},
title = {Online Ingarianti and Purwaningrum's Integrative Career Counseling Model: An Analysis of Implementation},
year = {2022},
isbn = {9781450386920},
publisher = {Association for Computing Machinery},
address = {New York, NY, USA},
url = {https://doi.org/10.1145/3516875.3516986},
doi = {10.1145/3516875.3516986},
booktitle = {Proceedings of the 5th International Conference on Learning Innovation and Quality Education},
articleno = {94},
numpages = {9},
location = {Surakarta, Indonesia},
series = {ICLIQE '21}
}

@article{doi:10.1177/2158244020941844,
author = {Patricia Mawusi Amos and P. K. A. Bedu-Addo and Theresa Antwi},
title ={Experiences of Online Counseling Among Undergraduates in Some Ghanaian Universities},

journal = {Sage Open},
volume = {10},
number = {3},
pages = {2158244020941844},
year = {2020},
doi = {10.1177/2158244020941844},

URL = { 
    
        https://doi.org/10.1177/2158244020941844
    
    

},
eprint = { 
    
        https://doi.org/10.1177/2158244020941844
    
    

}
}

@inproceedings{10.1145/1940761.1940772,
author = {Toyama, Kentaro},
title = {Technology as amplifier in international development},
year = {2011},
isbn = {9781450301213},
publisher = {Association for Computing Machinery},
address = {New York, NY, USA},
url = {https://doi.org/10.1145/1940761.1940772},
doi = {10.1145/1940761.1940772},
booktitle = {Proceedings of the 2011 IConference},
pages = {75–82},
numpages = {8},
location = {Seattle, Washington, USA},
series = {iConference '11}
}
\appendix
\update{
\section{Appendix A: Questionnaire}\label{appendix}
\begin{itemize}
    \item Questions on background and motivation:
    \begin{itemize}
        \item What kind of prior experience did you have in skill training?
        \item What motivated you to join the Livelihood College?
    \end{itemize}
    \item Awareness and mobilization:
    \begin{itemize}
        \item How do you make people aware of the facilities at the Livelihood College?
        \item How do you reach out to interior parts of the district?
        \item What happens in the counseling stage?
        \item What kind of support do you receive in mobilization? Who helps you in the process?
        \item What kind of challenges do you face in mobilizing and creating awareness?
        \item Why do you think trainees turn up in fewer numbers?
    \end{itemize}
    \item Training:
    \begin{itemize}
        \item What kind of challenge do you face in training?
        \item Resource availability:
        \item What kind of resources do you utilize to train the trainees?
        \item Did you ever face resource shortages? If yes, what kind?
    \end{itemize}
    \item Digital tool use:
    \begin{itemize}
        \item What are the steps involved in registering attendance using the Biometric system?
        \item How do you utilize the CCTV cameras?
    \end{itemize}
    \item Business persons
    \begin{itemize}
        \item Do you think the trainees from the Livelihood college are well trained? If No, what more is required? If Yes, can you explain how they add value to your enterprise?
        \item Is the training in line with on-the-job requirements? What more can be added?
    \end{itemize}
\end{itemize}
}
\end{document}